\documentclass[11pt]{article}

\usepackage[a4paper,margin=1in]{geometry}
\usepackage[english]{babel}

\usepackage[T1]{fontenc}
\usepackage[utf8]{inputenc}

\usepackage{microtype}
\usepackage{setspace}
\usepackage{amsmath,mathtools}
\usepackage{amsthm}
\usepackage{bm}
\usepackage{bbm}

\usepackage{newtxtext,newtxmath}
\theoremstyle{remark}

\usepackage{graphicx}
\usepackage{caption}
\usepackage{subcaption}
\usepackage{float}

\usepackage{booktabs}
\usepackage{multirow}
\usepackage{tabularx}
\usepackage[table]{xcolor}
\usepackage{threeparttable}
\usepackage{adjustbox}
\usepackage{siunitx}
\usepackage{enumitem}

\usepackage[ruled,vlined,noend]{algorithm2e}
\SetAlgorithmName{}{}{}

\SetAlgoCaptionSeparator{}

\usepackage[most]{tcolorbox}
\newtcolorbox{keyidea}{
  colback=gray!5,
  colframe=gray!60,
  boxrule=0.4pt,
  arc=2pt,
  left=6pt,right=6pt,top=4pt,bottom=4pt
}

\usepackage[authoryear,round]{natbib}
\bibpunct{(}{)}{;}{a}{,}{,}

\usepackage[hidelinks]{hyperref}

\usepackage{etoolbox}
\AtBeginEnvironment{thebibliography}{%
  \renewcommand{\url}[1]{}%
  \renewcommand{\href}[2]{#2}%
}

\newcommand{\cprior}{\varpi}

\newcommand{\keywords}[1]{%
  \par\bigskip\noindent\textbf{Keywords: }#1\par
}

\title{A Portfolio-Anchored Frequency–Severity Risk Index for Trip and Driver Assessment Using Telematics Signals}

\author{Jongtaek Lee, Andrei Badescu, X. Sheldon Lin\\[1ex]
Department of Statistical Sciences, University of Toronto}

\date{}

\begin{document}
\maketitle

\begin{abstract} In this paper, we propose a novel \emph{frequency–severity joint trip-level risk index} that combines the frequency of abnormal driving patterns with a severity component reflecting how extreme such behavior is relative to a portfolio-level baseline. Severity is quantified through an inverse-probability penalty that increases with the rarity of observed tail extremes, rather than being interpreted as a claim size. Based on high-frequency telematics data, we construct a multi-scale representation of longitudinal acceleration using the maximal overlap discrete wavelet transform (MODWT), which preserves localized driving patterns across multiple time scales. To capture severity as tail rarity, we model the portfolio distribution using a Gaussian–Uniform mixture with a layered tail structure, where Gaussian components describe typical driving behavior and the tail is partitioned into ordered severity layers that reflect increasing extremeness. We develop a likelihood-based estimation procedure that makes inference feasible for this mixture model. The resulting severity layers are then used to construct multi-layer tail counts (MLTC) at the trip level, which are modeled within a Poisson–Gamma framework to yield a closed-form posterior risk index that jointly reflects frequency and severity. This conjugate structure naturally supports sequential updating, enabling the construction of dynamically evolving driver-level risk profiles. Using the UAH-DriveSet controlled dataset, we demonstrate that the proposed index enables reliable discrimination across behavioral driving states, identification of high-risk trips, and coherent ranking of drivers, yielding a purely behavior-driven risk measure suitable for actuarial ratemaking and potentially mitigating fairness concerns associated with traditional covariates.
\end{abstract}

\bigskip

\keywords{Telematics signal, maximal overlap wavelet transform, multiple time scales, Gaussian-Uniform mixture model, MU-MEMR, tail layers, portfolio distribution, severity weights, Poisson--Gamma structure, frequency--severity joint trip-level risk index.}

\section{Introduction}
\label{sec:intro}
\par Traditional auto insurance ratemaking relies primarily on static driver and vehicle characteristics, which are typically evaluated at discrete points in time (e.g., annually). While these covariates are informative for assessing long-term or average risk levels, they are inherently limited in their ability to capture real-time driving behavior, as they are only indirectly related to the instantaneous risk of an accident, see \citet{chan2025assessing}, \citet{VerbelenAntonioClaeskens2018}. For the same reason, because these covariates are inherently attached to the driver or vehicle, pricing is typically conducted at the driver level and remains largely static over time, rather than being dynamically adjusted from trip to trip. With recent technological advances enabling high-frequency data collection, telematics has emerged as a means to address the limitations of traditional ratemaking. This shift supports pricing frameworks based directly on observed driving behavior, allowing insurers to adjust premiums dynamically through usage-based insurance (UBI), see \citet{VerbelenAntonioClaeskens2018}. In this context, telematics signals are used to refine risk prediction by directly measuring how a vehicle is operated at high temporal resolution. These signals can be incorporated into insurance practice through pricing, screening and underwriting, or ongoing monitoring. Their value is particularly pronounced in settings where claims are sparse, as signal-based approaches allow insurers to infer risk from observed driving behavior rather than relying solely on infrequent loss realizations. 

\par Telematics data comprise both kinematic signals (e.g., speed, longitudinal and lateral acceleration) and contextual information (e.g., road type, weather, traffic conditions, and time of day). In practice, however, such data rarely include ground-truth labels of accident risk at the trip or moment level. This absence of labeled outcomes motivates telematics research to focus on how effectively risk-relevant patterns can be represented and quantified directly from high-frequency signals, rather than relying only on traditional covariates, claims data, or explicit trip-level labels. 

\par The actuarial telematics literature has rapidly evolved, providing a variety of methods to manage this rich information for insurance applications. The first strand focuses on ``representation and feature construction'' to turn high-frequency driving records into actuarial covariates, including covariate selection and engineered summaries from telematics signals \citet{Wuthrich2017CovariateSelection}, low-dimensional representations such as speed--acceleration heatmaps and related feature extraction frameworks \citet{GaoWuthrich2018Heatmaps}, \citet{GaoMengWuthrich2019SAJ}, as well as broader data mining approaches that uncover common driving patterns from telematics records \citet{ChanTseungBadescuLin2025NAAJ}. The second strand develops ``actuarial modeling and pricing frameworks'' that incorporate telematics information into insurance models and rating systems, ranging from credibility-style multivariate formulations and model-based integration \citet{DenuitGuillenTrufin2019AAS}, to assessments of the incremental value of dynamic telematics updates \citet{HenckaertsAntonio2022IME}, bonus--malus style scoring for sequential ratemaking \citet{YanezGuillenNielsen2025BMS}, as well as empirical studies on how telematics improves claims prediction and classification \citet{Meng2022}, \citet{DuvalBoucherPigeon2022NAAJ}, \citet{DuvalBoucherPigeon2024CANN}. Many approaches in this strand essentially reduce high-frequency telematics time series to engineered features or summary statistics, but a large portion of the temporal structure in the signals has remained unexploited. This compression has limitations in detecting risk-relevant patterns that are localized in time.  This motivated the third strand ``driving dynamics and state evolution'' that leverages the temporal structure of the signals by using sequential or latent state models to describe how risk-relevant behavior unfolds over time and how deviations from typical behavior can be identified \citet{Meng2022}, \citet{JiangShi2024HMM}, \citet{chan2025assessing}. In particular, \citet{chan2025assessing} propose a multivariate Continuous-Time Hidden Markov Model (CTHMM) to model a trivariate response comprising speed, longitudinal acceleration, and lateral acceleration. They construct an anomaly index that quantifies the number of abnormal behavioral episodes observed during a trip. While this approach is effective in identifying aggressive driving patterns, it seems less successful in capturing drowsy or inattentive behavior.

\par Despite these methodological advances, most telematics-based risk assessments remain fundamentally \emph{frequency-oriented}. Existing approaches primarily emphasize \emph{how often} abnormal behavior occurs or the magnitude of deviations from a fitted model, with comparatively limited attention to \emph{how severe} those behaviors are. Within this trajectory toward increasingly signal-driven risk evaluation, incorporating behavioral \emph{severity} as an additional dimension represents a natural and necessary extension to achieve a more refined and comprehensive risk assessment framework.

\subsection{Data and Preliminary Analysis}
\par This section describes the data that will be used in this paper and presents a preliminary analysis to justify the introduction of telematics \emph{severity} to risk evaluation. The analysis will show how frequency-based summaries can miss some important risk-relevant patterns and how severity-based summaries can help a more accurate identification of those patterns, which motivates our \emph{frequency--severity joint risk index} that will be proposed in the following sections. This naturally raises the question of why severity information should provide a complementary and practically useful dimension beyond frequency, which we illustrate next via a simple empirical justification.

\subsubsection{Data}
\par The UAH-DriveSet introduced by \citet{Romera2016} is a public smartphone-sensor dataset for driver behavior analysis. In this paper, we use this Controlled Data which was recorded using the DriveSafe application. The dataset includes recordings from six different drivers with their own vehicles under three behavioral driving states: \texttt{normal}, \texttt{aggressive}, and \texttt{drowsy}. Normal driving reflects typical everyday behavior, aggressive driving shows impatience and abrupt maneuvers such as overspeeding or tailgating, and drowsy driving simulates mild sleepiness with occasional inattention. 

\par Data were collected on two pre-specified routes near Alcal\'a de Henares in the Community of Madrid, Spain: a secondary road route (about 90~km/h maximum speed) and a motorway route (about 120~km/h maximum allowed speed). Each driver repeats these routes under the three instructed driving states, yielding multiple trips per driver on each road type. On the secondary road, each of Drivers~1--5 contributes four trips (two \texttt{normal} repeats, one \texttt{aggressive}, and one \texttt{drowsy}), whereas on the motorway each driver contributes one trip per instructed state (\texttt{normal}, \texttt{aggressive}, \texttt{drowsy}). The only exception is Driver~6, whose electric vehicle completes the motorway set but provides only \texttt{normal} and \texttt{drowsy} on the secondary road due to limited vehicle autonomy. Consequently, there are 40 trips, and each operated over about 8 to 20 minutes. Table~\ref{tab:controlled_data_summary} summarizes the drivers, vehicles, and the number of trips by road type. In this paper, we solely focus on longitudinal acceleration recorded at 10~Hz (every 0.1 seconds) in units of $G$ ($9.8\,\mathrm{m/s^2}$), where positive values indicate acceleration and negative values indicate braking. This Controlled Data have also been analyzed in \citet{chan2025assessing}, which makes it a useful benchmark for contrasting different signal-based risk evaluation approaches.

\begin{table}[H]
\centering
\caption{Summary of the UAH-DRIVESET controlled dataset: driver demographics, vehicle information, and the number of available trips by road type (secondary vs.\ motorway).}
\label{tab:controlled_data_summary}
\footnotesize
\setlength{\tabcolsep}{4pt}
\renewcommand{\arraystretch}{1.05}
\begin{tabular}{l l l l l c c c}
\toprule
Driver & Gender & Age & Vehicle & Fuel & Sec.\ trips & Mot.\ trips & Total \\
\midrule
D1 & Male & 40--50 & Audi Q5 (2014)          & Diesel   & 4 & 3 & 7 \\
D2 & Male & 20--30 & Mercedes B180 (2013)    & Diesel   & 4 & 3 & 7 \\
D3 & Male & 20--30 & Citro\"en C4 (2015)     & Diesel   & 4 & 3 & 7 \\
D4 & Female & 30--40 & Kia Picanto (2004)      & Gasoline & 4 & 3 & 7 \\
D5 & Male & 30--40 & Opel Astra (2007)       & Gasoline & 4 & 3 & 7 \\
D6 & Male & 40--50 & Citro\"en C-Zero (2011) & Electric & 2 & 3 & 5 \\
\bottomrule
\end{tabular}
\end{table}

\subsubsection{Preliminary Analysis}
\par
\begin{figure}[H]
  \centering
  \makebox[\textwidth][c]{%
    \includegraphics[width=1.1\textwidth,height=0.40\textheight]{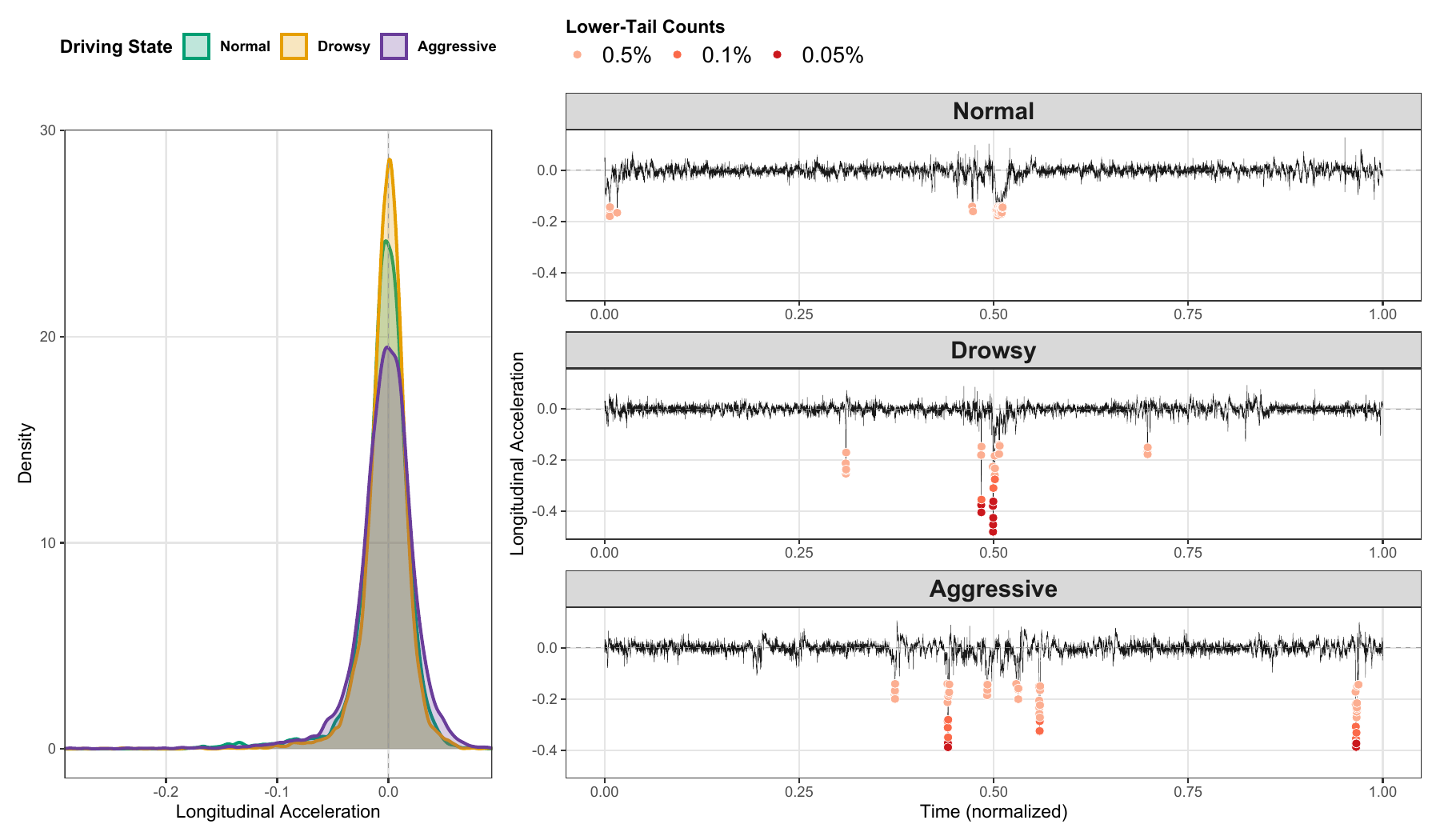}%
  }
  \caption{Driver~1 motorway example: longitudinal acceleration densities (left) and trip trajectories with pooled left-tail thresholds marked at the 0.5\%, 0.1\%, and 0.05\% quantiles (right), illustrating how tail observations are turned into MLTC events.}
  \label{fig:raw_traj_density}
\end{figure}

 For a comparison of the overall distributional profiles of different driving states, the empirical densities for longitudinal acceleration of Driver~1’s three motorway trips are displayed in Panel~(a) in Figure~\ref{fig:raw_traj_density}. In the panel, we can see that the normal and drowsy trips both concentrate near zero, and the drowsy trip places even more mass around zero than the normal trip. This illustrates the difficulty of identifying risk-relevant patterns of the drowsy or inattentive trip because it looks even safer than normal from a distributional perspective. In the following, we call this \emph{normal-looking profile}. This normal-looking profile implies that risk-relevant differences may be \emph{hidden} in the tails rather than in the bulk of the distribution, so there is a need to conduct a closer examination of the structural difference among the driving states of a driver by inspection of their tail behaviors.

\par We apply simple quantile layers to examine the tail structure. These quantile layers here are only a descriptive proxy to illustrate tail depth, while later sections define \emph{severity} formally using our proposed model. We compute the pooled lower tail quantiles based on the three trips and then commonly apply them to all the trips, as shown in Panel~(b) in Figure~\ref{fig:raw_traj_density}. Specifically, we combine all acceleration observations across the trips into one sample and compute its $0.5\%$, $0.1\%$, and $0.05\%$ lower tail quantiles, denoted by $q_{0.5\%}$, $q_{0.1\%}$, and $q_{0.05\%}$. Values below these thresholds represent increasingly severe deceleration-relevant patterns, and Table~\ref{tab:tail-exceedance-severity} reports the corresponding layer counts.

\par The result shows that multiple depths in the tails carry information that a frequency-based summary with a single threshold can miss. For example, if one only counts exceedances beyond $q_{0.5\%}$, the normal trip would appear riskier than the drowsy trip (sum of exceedances are $41$ versus $21+3+7=31$). In contrast, the deeper layers reveal that the drowsy trip places more mass in the extreme tail. The drowsy trip has $3$ exceedances in $(q_{0.05\%},q_{0.1\%}]$ and $7$ exceedances below $q_{0.05\%}$, while the normal trip has none in either layer, which makes it separable once the severity-based summary is taken into account. This mechanism is consistent with the frequency-based approach in \citet{chan2025assessing}, under which drowsy trips in the same Controlled Data were not flagged as risky by their CTHMM model. This preliminary analysis provides concrete evidence that incorporating the severity structure based on multiple tail-layers enables a more refined evaluation of driving behavior in telematics signals. This motivates our frequency--severity joint risk index to more accurately identify risky trips.

\begin{table}[H]
\centering
\caption{Driver~1 motorway example: counts of lower-tail observations in the pooled distribution, grouped into three severity layers defined by empirical left-tail quantiles (0.5\%, 0.1\%, and 0.05\%).}
\label{tab:tail-exceedance-severity}
\footnotesize
\begin{tabular}{lccc}
\toprule
Driving state 
& $\#\{q_{0.1\%} < \cdot \le q_{0.5\%}\}$ 
& $\#\{q_{0.05\%} < \cdot \le q_{0.1\%}\}$ 
& $\#\{\cdot \le q_{0.05\%}\}$ \\
\midrule
Normal      & 41 & 0  & 0 \\
Drowsy      & 21 & 3  & 7 \\
Aggressive  & 39 & 10 & 6 \\
\bottomrule
\end{tabular}
\end{table}

\subsection{Severity Layers and a Trip-Level Risk Index}
\par The primary objective of this paper is to construct a frequency–severity joint trip-level risk index from telematics signals and to demonstrate its effectiveness in identifying dangerous trips and high-risk drivers. We further show that the proposed index discriminates between normal and risky (i.e., aggressive and drowsy) driving within a classification framework. Very importantly, the framework does not rely on traditional covariates, claims data, or risk labels. Instead, it evaluates risk directly at the \emph{trip level} using telematics signals alone, which aligns with practical UBI settings where claims are sparse.

\par Two core contributions support this objective. First, we introduce a formal notion of \emph{severity} into telematics-based risk evaluation, extending beyond purely frequency-based approaches. By decomposing tail behavior into \emph{severity layers} and assigning layer-specific weights based on portfolio-level rarity, we obtain a more granular characterization of abnormal patterns. Second, we develop a \emph{Gaussian--Uniform mixture} severity model with \emph{multiple} Uniform components on each tail, together with an implementable fitting algorithm described in Section~\ref{sec:severity-model}.

\par In classical insurance terminology, “severity” refers to the loss amount conditional on a claim. Here, we reinterpret \emph{severity} as the magnitude of abnormal longitudinal acceleration, quantified relative to its statistical rarity in the portfolio-level distribution. Observations lying deeper in the tails receive greater weight in the proposed index. Intuitively, suppose that each telematics signal exhibits a portfolio-level range of typical behavior (e.g., ordinary braking or turning intensity), then patterns that fall substantially outside this range, such as rare harsh braking or turning episodes, should contribute more heavily to risk assessment. We formalize this idea through severity layers that assign increasing weights to progressively rarer behaviors, independent of claim occurrence.

\par The framework integrates three components.
\begin{itemize}[leftmargin=1.4em, itemsep=2pt, topsep=2pt]
\item \textbf{Signal representation via wavelet transform:} We represent each telematics signal using the Maximal Overlap Discrete Wavelet Transform (MODWT) \citet{percival2013wavelet} to capture localized patterns across multiple time scales. Coefficients are aggregated across scales into a single trip-level series summarizing multi-scale behavior, yielding a per-exposure index normalized for trip length.

\item \textbf{Portfolio-level severity model:} We construct the portfolio distribution of aggregated wavelet coefficients and fit a \emph{Gaussian--Uniform mixture} \citet{coretto2011mle}, \cite{Coretto2022}, where Gaussian components represent typical behavior and Uniform components define abnormal tail behavior. Estimating severity at the portfolio level ensures comparability across trips and drivers.

\item \textbf{Trip-level frequency model and risk index:} Layer-wise trip counts are modeled using a Poisson--Gamma structure. The proposed risk index is defined as a severity-weighted per-exposure risk rate across layers and can be updated sequentially to produce a dynamic driver-level profile.
\end{itemize}

\par The focus on longitudinal acceleration is deliberate, as prior telematics studies have established its relevance for detecting risky driving behavior \citet{chan2025assessing}, \citet{YanezGuillenNielsen2025BMS}, \citet{Mao2021DecisionAdjusted}, and it provides a clear and interpretable starting point. The framework, moreover, is signal-agnostic: it can be applied to other telematics responses by fitting the corresponding severity and frequency components, and it can be extended to a multivariate setting by combining signal-specific indices into a composite risk measure. Such extensions are left for future work.

\par The resulting index functions as an \emph{anomaly index} that jointly accounts for both the frequency and severity of abnormal behaviors within a trip. In insurance applications, it yields a per-exposure trip score that can be aggregated and updated sequentially to maintain an evolving driving risk profile for pricing and underwriting in UBI. In transportation safety and fleet management, trip-level scores can prioritize review and intervention by highlighting the most concerning patterns. More broadly, the framework provides a signal-based screening metric applicable even when claims or labeled outcomes are limited.

\par The remainder of the paper is organized as follows. Section~2 reviews the MODWT representation and across-level (-scale) aggregation. Section~3 introduces the Gaussian--Uniform mixture severity model and its estimation procedure. Section~4 develops the Poisson--Gamma frequency model and defines the trip- and driver-level frequency–severity indices. Section~5 presents the empirical application and classification results, and Section~6 concludes and discusses future directions.

\section{Discrete Wavelet Transform for Telematics Signals}
\label{sec:dwt}

\par In this section, we develop a trip-level representation that will serve as input to our proposed risk index. For each trip, we decompose the signal into wavelet coefficients across multiple time scales and then aggregate these coefficients into a single series that summarizes driving patterns over both short and long durations. The resulting aggregated series is used to form an input portfolio distribution, which is a key ingredient of the proposed risk index in the subsequent sections. As discussed in Section~\ref{sec:intro}, this representation applies to any telematics signal recorded as a time series and can be extended to a multivariate risk index that integrates multiple telematics signals in future work. 

\par To achieve this, we employ a \emph{wavelet transform} \citet{percival2013wavelet}, which is particularly well suited for detecting abnormal patterns in telematics signals across multiple time scales. Telematics signals typically exhibit abrupt, short-lived events (e.g., harsh braking or acceleration), making global or overly smooth representations less effective. For example, Fourier bases have global support over the entire observation window and B-spline bases impose smoothness. In contrast, wavelets are explicitly designed to capture localized features in time: wavelet bases are \emph{localized} in time and operate through filters of varying duration, from short to long scales, that concentrate around points of change. The resulting wavelet coefficients are therefore associated with specific time intervals and scales, enabling precise identification of transient behaviors. This localization property makes the wavelet transform a natural and effective representation for analyzing telematics signals and extracting risk-relevant driving patterns.

\par In this paper, we use the \emph{maximal overlap discrete wavelet transform} (MODWT) rather than a regular discrete wavelet transform (DWT) because it is a better choice for telematics signals when our goal is to detect abnormal driving behavior while retaining \emph{all} possible features that could later be identified as abnormal patterns. Since the regular DWT does downsampling, it critically depends on where we ``break into'' the series, i.e., what we take as a starting point or origin for analysis. In other words, the regular DWT is sensitive to shifts so its filter might not line up well with interesting features in a time series \citet{percival2013wavelet}. In contrast, as you may have already noticed by the term ``maximal overlap'', the MODWT applies the wavelet and scaling filters at \emph{every} time point through circular shifts, considering all possible placements of the filters with the signal. This prevents missing important driving behaviors that occur at arbitrary times within a trip because the MODWT produces coefficients at every time point without downsampling.

\par We now formalize this representation by defining the MODWT filters and the resulting wavelet and scaling coefficients for each trip. These coefficients will be aggregated to form the input portfolio distribution for severity and frequency modeling.

\subsection{MODWT Decomposition and Representation}
\subsubsection{MODWT Filters}
\par A regular DWT is defined by two filters: a \emph{wavelet} filter $\{h_\ell\}_{\ell=0}^{L-1}$ and a \emph{scaling} filter $\{g_\ell\}_{\ell=0}^{L-1}$, where $L$ is the number of terms of the filters. To effectively capture a common characteristic of telematics signals, we adopt the \emph{Daubechies D4} wavelet filter, with $L=4$ terms, given by:
\[
h_0=\frac{1-\sqrt{3}}{4\sqrt{2}}, \quad h_1=\frac{-3+\sqrt{3}}{4\sqrt{2}}, \quad h_2=\frac{3+\sqrt{3}}{4\sqrt{2}}, \quad h_3=\frac{-1-\sqrt{3}}{4\sqrt{2}}.
\]
The corresponding scaling filter is obtained by
\[
g_\ell = (-1)^{\ell+1}\,h_{L-1-\ell},\qquad \ell=0,1,2,3,
\]
that is,
\[
g_0=-h_3,\quad g_1=h_2,\quad g_2=-h_1,\quad g_3=h_0.
\]
Telematics signals often exhibit \emph{short-term mean reversion} characteristic: the signal typically moves back toward the mean zero over the next few time points after a burst, which makes Gaussian assumptions for some telematics signals (e.g., longitudinal or lateral accelerations) reasonable, as demonstrated in \citet{chan2025assessing}. Another widely used filter is the \emph{Haar} wavelet filter, that is essentially based on the difference between adjacent values ($h_0=\frac{1}{\sqrt{2}}$, $h_1=\frac{-1}{\sqrt{2}}$), so this structure can underestimate the strength of localized deviations due to local cancellation by averaging adjacent values away. In contrast, the D4 filter distributes weights over multiple time points, which aligns with the mean reversion characteristic. This avoids underestimating important features in the transform domain and helps retain the signature of risk-relevant patterns in mean-reverting signals. Thus, the D4 filter provides a clear representation of meaningful changes in telematics signals.

\par A key feature of the wavelet transform is its \emph{multi-resolution decomposition}: it represents a time series using components that capture changes over different time scales. We index these scales by a \emph{level} $j$, which corresponds to changes in the original time series on a scale of $2^{j-1}$. At finer levels (small $j$), the transform captures short-lived, localized changes, while at coarser levels (large $j$), it captures smoother changes that persist over longer durations. To account for the multi-scale view, we denote by $\{h_{j,\ell}\}_{\ell=0}^{L_j-1}$ and $\{g_{j,\ell}\}_{\ell=0}^{L_j-1}$ the filters at level $j$ obtained from the DWT filters $\{h_\ell\}$ and $\{g_\ell\}$ by dyadic dilation (i.e., by inserting $2^{j-1}-1$ zeros between successive terms). For the MODWT, we use rescaled filters defined as
\begin{equation}
    \label{eq:MODWT filters}
    \widetilde{h}_{j,\ell} = \frac{h_{j,\ell}}{\sqrt{2}},\qquad
    \widetilde{g}_{j,\ell} = \frac{g_{j,\ell}}{\sqrt{2}},\qquad \ell=0,\ldots,L_j-1,
\end{equation}
where the filter width at level $j$ is
\begin{equation}
\label{eq:filter_width}
    L_j = (2^j-1)(L-1)+1,    
\end{equation}
i.e., the number of lags of the original signal references when computing each MODWT coefficient at level $j$. The rescaling ensures the variance contributions across levels are comparable. Without scaling, the contribution would be inflated simply because the MODWT produces coefficients at every time point.

\subsubsection{MODWT Coefficients}
\par To represent a telematics signal across time scales, we apply MODWT filters and define its \emph{wavelet} and \emph{scaling coefficients} decomposed at each level $j$. For trip $i$, let the telematics signal have the length-$T_i$ time series
\[
\mathbf{X}_i := (X_{i,t})_{t=0}^{T_i-1}.
\]
Given the MODWT filters in \eqref{eq:MODWT filters}, the corresponding coefficients at level $j$ are defined by
\begin{equation}
\label{eq:modwt-coef-def}
\widetilde{W}_{i,j,t} := \sum_{\ell=0}^{L_j-1}\widetilde{h}_{j,\ell}\,X_{i,(t-\ell)\bmod T_i},
\qquad
\widetilde{V}_{i,j,t} := \sum_{\ell=0}^{L_j-1}\widetilde{g}_{j,\ell}\,X_{i,(t-\ell)\bmod T_i},
\qquad t=0,\ldots,T_i-1,
\end{equation}
where $(t-\ell)\bmod T_i$ denotes circular indexing that maps any integer to an element of $\{0,1,\ldots,T_i-1\}$. For any integer $J\ge 1$, the vectors have the following interpretations:
\begin{itemize}
    \item a vector $\widetilde{\mathbf{W}}_{i,j}$ contains the MODWT wavelet coefficients associated with changes in $\mathbf{X}_i$ on a scale of $2^{j-1}$ for $j = 1, \dots, J$,
    \item and a vector $\widetilde{\mathbf{V}}_{i,J}$ contains the MODWT scaling coefficients associated with changes in $\mathbf{X}_i$ on a  scale $2^{J}$ or higher,
\end{itemize}
where $\widetilde{\mathbf{W}}_{i,j} = (\widetilde{W}_{i,j,t})_{t=0}^{T_i-1}$ and
$\widetilde{\mathbf{V}}_{i,J} = (\widetilde{V}_{i,J,t})_{t=0}^{T_i-1}$ \citet{percival2013wavelet} and $J$ is the chosen maximum level of the MODWT decomposition. Since the filter width $L_j$ increases with $j$, the coefficients at finer levels represent localized features and coarser levels represent milder features with longer duration. This multi-resolution structure across levels can capture both brief events and sustained patterns that can be informative driving behaviors.

\par The wavelet filters and coefficients in \eqref{eq:modwt-coef-def} provide a complete representation of the original time series $\mathbf{X}_i$. For each level $j$, we extend the filters $\{\widetilde{h}_{j,\ell}\}_{\ell=0}^{L_j-1}$ and $\{\widetilde{g}_{j,\ell}\}_{\ell=0}^{L_j-1}$ to length $T_i$ by setting $\widetilde{h}_{j,\ell}=0$ and $\widetilde{g}_{j,\ell}=0$ for $\ell=L_j,\ldots,T_i-1$. Then, we represent $\mathbf{X}_i$ using
\begin{equation}
\label{eq:representation of X}
    X_{i,t} \;=\; \sum_{j=1}^{J}\ \sum_{\ell=0}^{T_i-1}\widetilde{h}_{j,\ell}\,\widetilde{W}_{i,j,(t+\ell)\bmod T_i} \;+\; \sum_{\ell=0}^{T_i-1}\widetilde{g}_{J,\ell}\,\widetilde{V}_{i,J,(t+\ell)\bmod T_i},
\end{equation}
for a given $J\ge1$. Under the circular indexing convention in \eqref{eq:modwt-coef-def}, this reconstruction holds for all $t = 0, \dots, T_i-1$, so $\mathbf{X}_i$ can be recovered from $\{\widetilde{\mathbf{W}}_{i,j}\}_{j=1}^{J}$ and $\widetilde{\mathbf{V}}_{i,J}$. This wavelet transform is particularly useful in UBI settings where we need to focus on signal-based analysis rather than sparse claim counts because this representation allows us to map each trip to a coefficient series that retains the full signal information, which will serve as the foundation for our risk indices.

\subsection{Energy Decomposition and Aggregation}
\subsubsection{Energy and Variance Contributions}
\par As the level $j$ increases, coefficients beyond a certain level tend to represent only broad baseline movement, which is around zero, rather than meaningful risk-relevant maneuvers in a telematics signal. For this reason, choosing the maximum level of the decomposition $J$ is a natural question to select a meaningful set of levels, and we therefore quantify the \emph{contribution} of each level $j$ to trip $i$ using the MODWT energy decomposition. For a fixed integer $J\ge1$, \citet{percival2013wavelet} proved that the MODWT yields the \emph{energy decomposition} 
\begin{equation}
\label{eq:energy decomposition}
\|\mathbf{X}_i\|^2 \;=\; \sum_{j=1}^{J}\|\widetilde{\mathbf{W}}_{i,j}\|^2 \;+\; \|\widetilde{\mathbf{V}}_{i,J}\|^2,
\end{equation}
because we have
\[
    \|\widetilde{\mathbf{W}}_{i,j}\|^2 + \|\widetilde{\mathbf{V}}_{i,j}\|^2 = \|\widetilde{\mathbf{V}}_{i,j-1}\|^2, \qquad j=1, 2, \dots
\]
with the definition $\widetilde{\mathbf{V}}_{i,0}:=\mathbf{X}_i$ and $\|\cdot\|$ denotes the $\ell_2$ norm.

\par To quantify contributions for each level, we define a \emph{sample variance} of $\mathbf{X}_i$ by
\[
\widehat{\sigma}^2_{\mathbf{X}_i} \;:=\; \frac{1}{T_i}\|\mathbf{X}_i\|^2-\overline{X}_i^{\,2},
\]
where $\overline{X}_i := T_i^{-1}\sum_{t=0}^{T_i-1} X_{it}$. For each level $j$, define the energy component
\[
\mathcal{V}_i(j) \;:=\; \frac{1}{T_i}\bigl\|\widetilde{\mathbf{W}}_{i,j}\bigr\|^2,
\]
yielding the \emph{variance contribution} given by
\begin{equation}
\label{eq:energy-share}
\rho_{i,j} \;:=\; \frac{\mathcal{V}_i(j)}{\widehat{\sigma}^2_{\mathbf{X}_i}}, \qquad j=1,\ldots,J.
\end{equation}
where $\rho_{i,j}\in[0,1]$ and the cumulative sum typically increases as $J$ increases. This variance contribution $\rho_{i,j}$ will be used for a practical choice of the decomposition depth $J$ in Section~\ref{sec:application}. In particular, since the incremental contribution $\rho_{i,j}$ typically decreases as $j$ increases, we can identify a cutoff level beyond which additional scales contribute negligibly to the overall variance and are therefore treated as practically insignificant for the representation.

\subsubsection{Across-level Aggregation}
\par Risk-relevant patterns in telematics signals do not occur in a single time scale within a trip, so a trip-level representation should capture both short- and long-duration features. We therefore aggregate MODWT wavelet coefficients $\widetilde{\mathbf{W}}_{i,j}$ across levels into a single time series which summarizes multi-scale patterns. In addition, this across-level aggregation enables a per-exposure risk index normalized for the trip length via an exposure offset, as will be described in Section~\ref{sec:frequency-model}.

\par For a set of levels $\mathcal{J}\subset\{1,\ldots,J\}$, we define the \emph{aggregated wavelet coefficient} for a trip $i$ at each time point $t$ as
\begin{equation}
\label{eq:aggregation}
    C_{i,t} \;:=\; \mathfrak{g} \Bigl(\{\widetilde{W}_{i,j,t}\}_{j\in\mathcal{J}}\Bigr), \qquad t=0,\ldots,T_i-1,
\end{equation}
where $\{\widetilde{W}_{i,j,t}\}_{j\in\mathcal{J}}$ is the collection of the MODWT coefficients across levels at time point $t$ and $\mathfrak{g}: \mathbb{R}^{|\mathcal{J}|}\to\mathbb{R}$ is the aggregation function. There are some practical choices of $\mathfrak{g}(\cdot)$ based on how the coefficients are aggregated across levels:
\begin{itemize}
    \item a maximum choice $C_{i,t}=\max_{j\in\mathcal{J}}|\widetilde{W}_{i,j,t}|$,
    \item an average pooling $C_{i,t}=|\mathcal{J}|^{-1}\sum_{j\in\mathcal{J}}|\widetilde{W}_{i,j,t}|$,
    \item a weighted pooling $C_{i,t}=\sum_{j\in\mathcal{J}}\omega_j\,|\widetilde{W}_{i,j,t}|$ with weights $\omega_j\ge0$ and $\sum_{j\in\mathcal{J}}\omega_j=1$,
    \item a single-level choice $C_{i,t}=\widetilde{W}_{i,j^*,t}$ for a fixed $j^*\in\mathcal{J}$.
\end{itemize}
The choice of the aggregation rule depends on the nature of data and the goal of the analysis.

\par This aggregation produces a single time series of the aggregated wavelet coefficients for trip $i$
\begin{equation}
\label{eq:agg_wav_coeff}
    \mathbf{C}_i=(C_{i,0},\ldots,C_{i,T_i-1}),
\end{equation}
which summarizes $\{\widetilde{W}_{i,j,t}\}_{j \in \mathcal{J}}$ across $J$ levels at each time point. Regardless of the choice of $\mathfrak{g}(\cdot)$, $\mathbf{C}_i$ is designed to aggregate driving behaviors that may appear at different temporal scales into a time series. In Section~\ref{sec:severity-model} and Section~\ref{sec:frequency-model}, $\mathbf{C}_i$ will be an ingredient to build a portfolio distribution that is the input for our proposed framework.

\section{Severity Model}
\label{sec:severity-model}

\par To define severity at the portfolio level for trip-level risk evaluation, we need a stable baseline that is not overly sensitive to individual trips or dominant driving styles in the portfolio. In this section, we therefore model the portfolio distribution using a severity model that absorbs tail behavior while preserving a stable representation of typical driving. The estimated severity structure will be used in Section~\ref{sec:frequency-model} to build the joint frequency--severity risk index.

\subsection{Motivation and General Description of the Model}
\par Longitudinal acceleration is often modeled as Gaussian, \citet{chan2025assessing}, due to its short-term mean reversion as discussed in Section~\ref{sec:dwt}; however, a pure Gaussian fit can be distorted by tail values that inflate the fitted variance. Moreover, residual-based outlier measures relative to a fitted model proposed by \citet{chan2025assessing} can be sensitive to how the ``normal'' reference is defined, because they implicitly assume the fitted model as normal driving, which may be biased when the training portfolio is dominated by a particular driving style. In this paper, we adopt a \emph{Gaussian--Uniform mixture model} \citet{coretto2011mle} to separate the \emph{normal} baseline from \emph{abnormal} tail behavior. The Gaussian component is dedicated to representing typical driving, whereas Uniform components absorb tail observations. Consequently, the Gaussian component remains stable against variance inflation, and the baseline used for severity modeling becomes less dependent on portfolio composition or some particular driving styles. This yields a more robust reference distribution for the severity structure.

\par Uniform endpoints make the mixture likelihood discontinuous, so \citet{Coretto2022} proposed Multiple Expectation–Maximization Runs (MEMR) to make fitting feasible for a model with a single Uniform component. Nevertheless, EM fitting still remains difficult and unsolved under \emph{multiple} Uniform components. In this paper, we extend the MEMR to multiple Uniform components and propose a \emph{Multi-Uniform MEMR} (MU-MEMR) algorithm, where each Uniform component defines its \emph{own} severity layer. Uniform components assigned to deeper tail intervals represent higher severity, while those closer to the Gaussian core represent lower severity. This novel extension is central to our severity framework because it enables a direct quantification of severity in terms of rarity in driving patterns. 

\par We define a \emph{portfolio distribution} of aggregated wavelet coefficients \eqref{eq:agg_wav_coeff} as the input to the Gaussian--Uniform severity model. The mixture model provides robustness conditional on the chosen portfolio sample by isolating typical behavior from tail behavior; however, the estimated baseline can still be affected by how the coefficients are pooled across trips. One consideration here is that trips may include behavior-irrelevant noise, so naive pooling can contaminate the baseline. For example, rough road surfaces can cause suspicious deviations not driven by the driver's maneuver, and long stop periods generate many near-zero coefficients for a trip. The other consideration is that aggregated coefficients within a trip form a time series, so naively pooling all coefficients can also violate the i.i.d.\ sampling assumption underlying likelihood fitting. We therefore construct the portfolio distribution using a balanced sampling scheme that controls both the \emph{representativeness} across trips and \emph{temporal dependence} structure within trips.

\par Specifically, we propose a \emph{hierarchical sampling with within-trip thinning} scheme that addresses both considerations. The hierarchical stage balances contributions by drawing (i) a driver uniformly and (ii) a trip from that driver uniformly. The thinning stage then reduces temporal dependence by retaining coefficients within each selected (or available) trip at time points spaced far enough apart so that the empirical autocorrelation falls below a chosen threshold. Pooling these retained coefficients across trips yields the portfolio sample $\mathbf{C}$, which we treat as an approximately i.i.d.\ input when defining and fitting the Gaussian--Uniform severity model in the following sections.
 
\subsection{Likelihood and Parameter Estimation}
\par Recall the aggregated wavelet coefficient series $\mathbf{C}_i=\{C_{it}\}_{t=0}^{T_i-1}$ for each trip $i$, defined in Section~\ref{sec:dwt}. To fit the severity model, we construct a portfolio sample $\mathbf{C}$ using the sampling scheme discussed above. Let $\mathcal{T}_i \subset \{0,\ldots,T_i-1\}$ denote the set of \emph{retained} time points after thinning for a selected trip $i$. We then form the portfolio sample by pooling the retained coefficients,
\[
\mathbf{C}:=\{C_r\}_{r=1}^n
= \bigcup_{i}\{C_{it}: t\in\mathcal{T}_i\},
\qquad n=\sum_i |\mathcal{T}_i|,
\]
where $r$ is a re-indexing of the retained $(i,t)$ pairs. Then, we model the portfolio sample $\mathbf{C}$ as an approximately i.i.d.\ sample from a Gaussian--Uniform mixture model.

\par For $g = 1, \dots, G$, define a family of Gaussian densities $\phi\bigl(c_r; \theta_g\bigr)$ with full support on $\mathbb{R}$, where $\theta_g = (\mu_g, \sigma_g)\in\mathbb{R}\times(0, +\infty)$ represents the mean and standard deviation, respectively. Let $m = 1, \dots, M^+$ index a collection of Uniform densities $u\bigl(c_r; \theta^+_m\bigr)$ on the right tail of the mixture model, where $\theta^+_m = (u^+_m, u^+_{m+1}) \in \mathbb{R}^2$ denotes the uniform support. Let $m' = 1, \dots, M^-$ index Uniform densities $u\bigl(c_r; \theta^-_{m'}\bigr)$ on the left tail, with the support $\theta^-_{m'} = (u^-_{m'+1}, u^-_{m'})\in\mathbb{R}^2$. Denote the membership probabilities by $\pi_g$ for each Gaussian, $\pi^+_m$ for each right-tail Uniform, and $\pi^-_{m'}$ for each left-tail Uniform component. These satisfy $\pi^-_{m'},\pi_g,\pi^+_m\in (0,1)$ and
\[
\sum_{m'=1}^{M^-}\pi^-_{m'}+\sum_{g=1}^G \pi_g+\sum_{m=1}^{M^+}\pi^+_{m}=1.
\]

\par Let $Z_r$ be a latent variable that labels which component generates $C_r$. The support of $Z_r$ consists of $M^-$ left-tail Uniform components, $G$ Gaussian components, and $M^+$ right-tail Uniform components. The prior probabilities are $\Pr(Z_r = m') = \pi^-_{m'}, \Pr(Z_r = g) = \pi_g, \Pr(Z_r = m) = \pi^+_m,$ and conditionally on $Z_r$ we have
\[
C_r \mid Z_r = m' \sim u\!\left(\bullet\,;\,\theta^-_{m'}\right),\quad
C_r \mid Z_r = g \sim \phi\!\left(\bullet\,;\,\theta_{g}\right),\quad
C_r \mid Z_r = m \sim u\!\left(\bullet\,;\,\theta^+_{m}\right).
\]
The resulting finite mixture density is
\begin{equation}
  f(\bullet\,;\,\eta)
  \;:=\;
  \sum_{m'=1}^{M^-}\pi^-_{m'}\,u\!\left(\bullet\,;\,\theta^-_{m'}\right)
  \;+\;
  \sum_{g=1}^{G}\pi_g\,\phi\!\left(\bullet\,;\,\theta_g\right)
  \;+\;
  \sum_{m=1}^{M^+}\pi^+_{m}\,u\!\left(\bullet\,;\,\theta^+_{m}\right).
\end{equation}
where $\eta=(\boldsymbol{\pi},\boldsymbol{\Theta})$ is the model parameter vector with $\boldsymbol{\pi}=\bigl(\{{\pi^-_{m'}}\}^{M^-}_{m'=1},\,\{{\pi_g}\}^{G}_{g=1},\,\{{\pi^+}_{m}\}^{M^+}_{m=1}\bigr)$ and $\mathbf{\Theta}=\bigl(\{\theta^-_{m'}\}_{m'=1}^{M^-},\,\{\theta_g\}_{g=1}^{G},\,\{\theta^+_{m}\}_{m=1}^{M^+}\bigr)$.

\par Given a realization $\mathbf{c} = \{c_1, c_2, \dots, c_n\}$ of the portfolio sample $\mathbf{C}$, the log-likelihood function is
\[
l_n(\eta)\;=\;\sum_{r=1}^{n} \log f\bigl(c_r;\,\eta\bigr).
\]
At iteration $h = 0, 1, 2, \dots$, the posterior probabilities for each data point $c_r$ and each component are
\[
\tau_{m'}(c_r;\eta^{(h)}) \;:=\; \Pr(Z_r=m'\mid C_r=c_r,\eta^{(h)}) \;=\; \frac{\pi_{m'}^{-(h)}\,u\bigl(c_r;\,\theta_{m'}^{-(h)}\bigr)}{f\bigl(c_r;\,\eta^{(h)}\bigr)},
    \quad 
    m'=1,\dots,M^-,
\]
\[
\tau_{g}(c_r;\eta^{(h)}) \;:=\; \Pr(Z_r=g\mid C_r=c_r,\eta^{(h)}) \;=\; \frac{\pi_g^{(h)}\,\phi\bigl(c_r;\,\theta_{g}^{(h)}\bigr)}{f\bigl(c_r;\,\eta^{(h)}\bigr)},
    \quad 
    g=1,\dots,G,
\]
\[
\tau_{m}(c_r;\eta^{(h)}) \;:=\; \Pr(Z_r=m\mid C_r=c_r,\eta^{(h)}) \;=\; \frac{\pi_{m}^{+(h)}\,u\bigl(c_r;\,\theta_{m}^{+(h)}\bigr)}{f\bigl(c_r;\,\eta^{(h)}\bigr)},
    \quad 
    m=1,\dots,M^+.
\]
The EM formulation leads to
\begin{equation*}
\begin{aligned}
Q(\eta; \eta^{(h)}) &= \sum_{r=1}^{n} \sum_{m'=1}^{M^-} \tau_{m'}(c_r; \eta^{(h)}) \log u(c_r; \theta^-_{m'}) \\
&\quad+ \sum_{r=1}^{n}\sum_{g=1}^{G} \tau_{g}(c_r; \eta^{(h)}) \log \phi(c_r; \theta_g) \\
&\quad + \sum_{r=1}^{n}\sum_{m=1}^{M^+} \tau_{m}(c_r; \eta^{(h)}) \log u(c_r; \theta^+_{m}) \\
&\quad + \sum_{r=1}^{n} \Bigg[ \sum_{m'=1}^{M^-} \tau_{m'}(c_r; \eta^{(h)}) \log \pi^-_{m'} 
+ \sum_{g=1}^{G} \tau_{g}(c_r; \eta^{(h)}) \log \pi_g + \sum_{m=1}^{M^+} \tau_{m}(c_r; \eta^{(h)}) \log \pi_m^+ \Bigg],
\end{aligned}
\end{equation*}
which decomposes accordingly:
\[
Q\bigl(\eta;\,\eta^{(h)}\bigr)
    \;=\;
Q_{u^-}\bigl(\theta_{u^-},\,\eta^{(h)}\bigr)
\;+\;
Q_{\phi}\bigl(\theta_{\phi},\,\eta^{(h)}\bigr)
\;+\;
Q_{u^+}\bigl(\theta_{u^+},\,\eta^{(h)}\bigr)
\;+\;
Q_{\pi}\bigl(\theta_{\pi},\,\eta^{(h)}\bigr).
\]

\subsection{Multi-Uniform MEMR Algorithm}
\par We now describe the fitting algorithm for the proposed MU-MEMR estimation procedure. The key difference from MEMR is the structure of tail layers. We split each tail into equally spaced partitions with resolutions $q$ and $p$, and, at each iteration, we select $M^-$ and $M^+$ tail layers from these partitions. We then \emph{reinitialize} the Uniform parameters using the selected layers and run EM for the Gaussian components and membership probabilities. This \emph{candidate search} makes EM fitting feasible for a Gaussian--Uniform mixture with multiple Uniform components.

\par For a robust initialization for the Gaussian components, we start with \emph{trimmed-k-means} \citet{Cuesta-Albertos1997}. The parameters are not affected by extreme values because it clusters the central bulks after trimming the most extreme $\alpha\%$ observations in $\mathbf{C}$. This also naturally provides the \emph{base grid} for the candidate search for Uniform parameters, given by the trimmed observations assigned to each tail. Let
\[
\mathbf{c}^- 
  = 
  \bigl\{c_1^-, c_2^-, \dots\bigr\},
\qquad
\mathbf{c}^+
  =
  \bigl\{c_1^+, c_2^+, \dots\bigr\},
\]
denote the trimmed observations assigned to the left and right tails, respectively. Define
\[
c^-_{\min} = \min(\mathbf{c}^-),\ c^-_{\max} = \max(\mathbf{c}^-),\ c^+_{\min} = \min(\mathbf{c}^+),\ c^+_{\max} = \max(\mathbf{c}^+),
\]
with $c^-_{\max}<0<c^+_{\min}$ by construction of the left and right tails. We split $[c^-_{\min}, c^-_{\max}]$ and $[c^+_{\min}, c^+_{\max}]$ into $q$ and $p$ equally spaced partitions, respectively, and set each split point to the nearest observed value in $\mathbf{c}^-$ and $\mathbf{c}^+$ when no observation lies exactly on it. This yields ordered points, $b^-_{(\cdot)}$, $b^+_{(\cdot)}$:
\[
c^-_{\min} < b^-_{(1)} < b^-_{(2)} < \cdots < b^-_{(q)} \le c^-_{\max} < 0,
\qquad
0 < c^+_{\min} \le b^+_{(1)} < b^+_{(2)} < \cdots < b^+_{(p)} < c^+_{\max}.
\]
We define the base grid for Uniform components as
\[
\mathcal{U}^- \;=\; \bigl\{b^-_{(1)},\dots,b^-_{(q)}\bigr\},
\qquad
\mathcal{U}^+ \;=\; \bigl\{b^+_{(1)},\dots,b^+_{(p)}\bigr\},
\]
where $\mathcal{U} = \mathcal{U}^- \cup \mathcal{U}^+$. For given $(M^-,M^+)$, each MU-MEMR run restricts the Uniform endpoints
\[
\bigl(u^-_{M^-+1},\dots,u^-_1,u^+_1,\dots,u^+_{M^++1}\bigr)
\]
to build a \emph{candidate} set. Then, we repeat the initialization for Uniform components within the set by selecting $\binom{q}{M^-}$ left-tail internal endpoints from the left base grid $\mathcal{U}^-$ and $\binom{p}{M^+}$ right-tail internal endpoints from the right base grid $\mathcal{U}^+$. Increasing $q$ and $p$ refines the resolution of tail-layers and yields a sieve-MLE view \citet{Grenander1981}, \citet{GemanHwang1982} because the base grid then yields an increasingly fine approximation to the continuous parameter space of all possible endpoints. 

\par To obtain stable estimation and an interpretable severity structure, we restrict the Gaussian--Uniform mixture parameters to a space defined by four constraints. One constraint controls component scales to avoid degenerate fits, and the other three define the severity structure by separating tails from the Gaussian core, covering the full tail range with consecutive Uniform layers, and enforcing monotone membership probabilities for Uniform components.

\par First, we impose a \emph{scale constraint} to both Gaussian and Uniform components. \citet{Day1969} demonstrated the unboundedness of the likelihood function of a Gaussian mixture when some scale parameter $\sigma_g \to 0$ for a fixed mean $\mu_g$. An analogous phenomenon occurs for Uniform components if $u_{m+1} \to u_m,$ for a fixed $u_m$ \citet{Tanaka2006}. To ensure the consistency of the MLE, we restrict a lower bound $\exp(-n^d) \le v_j$ for all $j$, where
\[
v_j = \begin{cases} 
{\dfrac{u^-_{m'} - u^-_{m'+1}}{\sqrt{12}}} & \text{if } j \text{ indexes } m' = 1, \dots, M^-, \\[6pt]
\sigma_g & \text{if } j \text{ indexes } g = 1,\dots,G, \\[6pt]
{\dfrac{u^+_{m+1} - u^+_{m}}{\sqrt{12}}} & \text{if } j \text{ indexes } m = 1, \dots, M^+,
\end{cases}
\]
with $n$ the sample size of $\mathbf{c}$ and $d \in (0,1)$ \citet{Tanaka2006}.

\par Second, we impose a \emph{tail-separation constraint} \citet{Coretto2022} to keep the Uniform tails separated from the Gaussian cores. We require
\[
u^-_{1} \;\le\; \mu_{1} - \delta \sigma_{1},
\qquad
\mu_{G} + \delta \sigma_{G} \;\le\; u^+_{1},
\]
where $\mu_1 \le \cdots \le \mu_G$ and $\delta>0$ is fixed. Here $\delta$ controls the overlap between the Gaussian cores and the Uniform tails.

\par Third, we impose a \emph{tail coverage constraint}. We fix the extreme tail endpoints as $u^-_{M^-+1}=c^-_{\min}$ and $u^+_{M^++1}=c^+_{\max}$, treating them as fixed outer parameters of the outermost Uniform components. We then set the Uniform supports to be consecutive intervals that share boundaries:
\[
\theta^-_{m'} = (u^-_{m'+1},\,u^-_{m'}), \quad m'=1,\dots,M^-, \qquad \theta^+_{m} = (u^+_{m},\,u^+_{m+1}), \quad m=1,\dots,M^+. 
\]
This constraint covers the full tail range and assigns every tail observation to exactly one Uniform component.

\par Fourth, we impose a \emph{monotonicity constraint} on Uniform membership probabilities for each tail so that deeper layers represent higher severity. With tail layers ordered from least extreme to most extreme, we require
\[
\pi^-_{1} \ge \pi^-_{2} \ge \cdots \ge \pi^-_{M^-},
\qquad
\pi^+_{1} \ge \pi^+_{2} \ge \cdots \ge \pi^+_{M^+}.
\]
We enforce these inequalities using \emph{isotonic regression} \citet{RobertsonWrightDykstra1988}. In each M-step, isotonic regression projects the updated probabilities onto the set of non-increasing sequences, while the ones for Gaussian components remain unchanged.

\par Then, we seek the maximizer of the log-likelihood $l_n(\eta)$ over the
following constrained parameter space:
\[
\eta_n := \left\{ \eta \in \boldsymbol{\eta} \;\middle|\;
\begin{aligned}
& v_j \ge \exp(-n^d) \quad \text{for all } j, \\
& u^-_{1} \le \mu_{1} - \delta \sigma_{1}, \quad \mu_{G} + \delta \sigma_{G} \le u^+_{1}, \\
& c^-_{\min} = u^-_{M^-+1}, \quad c^+_{\max} = u^+_{M^++1}, \\
& \theta^-_{m'}=(u^-_{m'+1},u^-_{m'}),\ \theta^+_{m}=(u^+_{m},u^+_{m+1}) \quad \text{for all } m',\;m, \\
& \pi^-_1 \ge \pi^-_2 \ge \cdots \ge \pi^-_{M^-}, \quad \pi^+_1 \ge \pi^+_2 \ge \cdots \ge \pi^+_{M^+}
\end{aligned}
\right\},
\]
with $d\in(0,1)$. The constrained maximum-likelihood estimator is then
\[
\hat{\eta}_n = \arg\max_{\eta\in\eta_n} l_n(\eta).
\]

\par We now introduce the MU-MEMR algorithm that implements this constrained
optimization of the Gaussian--Uniform mixture framework.

\begin{algorithm}[H]
\caption{\textbf{Multi-Uniform MEMR (MU-MEMR) Algorithm}}
\label{alg:mu_memr}

\KwIn{$\mathbf{c},\; d\in(0,1),\;\alpha\in(0,1)$}

\textbf{Initialize} $\theta_g,\;\theta_{\pi}$:\\
Given a fixed $G$, run \textit{trimmed-$k$-means} on the dataset $\mathbf{c}.$ 
\[
\mu^{(h=0)}_g \gets o_g,
\quad
\sigma^{(h=0)}_g \gets \max\{v_g,\exp(-n^d)\},
\quad
g=1,\dots,G,
\]
where $(o_g,\,v_g)$ are the means and standard deviations of the $G$ clusters. For fixed $M^-,M^+$,
\[
\pi^{-(h=0)}_{m'},\;\pi^{+(h=0)}_{m}
    \;\gets\;
    \frac{\alpha}{\,M^-+M^+\,},
    \quad
    m'=1,\dots,M^-,\quad m=1,\dots,M^+,
\]
\[
\pi^{(h=0)}_g
    \;\gets\;
    \frac{1-\alpha}{\,G\,},
    \quad
    g=1,\dots,G.
\]

\For{each choice of index sets 
$1 \le \Delta^-_1 < \cdots < \Delta^-_{M^-} \le q$ and $1 \le \Delta^+_1 < \cdots < \Delta^+_{M^+} \le p$ such that
$ b^-_{(\Delta^-_{m'+1})} - b^-_{(\Delta^-_{m'})} \ge \sqrt{12}\exp(-n^d)$ and 
$ b^+_{(\Delta^+_{m+1})} - b^+_{(\Delta^+_{m})} \ge \sqrt{12}\exp(-n^d)$ for all $m,m'$}
{

\textbf{Initialize} $s$-th $\theta^-_{m'},\;\theta^+_{m}$:
\[
u^{-(h=0)}_{M^-+1-m'}
    \gets
    b^-_{(\Delta^-_{m'})}, \quad u^{+(h=0)}_{m}
    \gets
    b^+_{(\Delta^+_{m})},
    \quad
    m'=1,\dots, M^-, \quad m=1,\dots,M^+,
\]
where $r$ indexes the combination of $(\Delta^-_1,\dots,\Delta^-_{M^-})$ and $(\Delta^+_1,\dots,\Delta^+_{M^+})$, and \[u^-_{M^-+1} \gets c^-_{\min}, \quad u^+_{M^++1} \gets c^+_{\min}
\] are fixed.\\[4pt]

Set
\(
\eta^{(h=0)}\gets
\bigl(\theta_{m'}^{-(h=0)},\,\theta_{g}^{(h=0)},\,\theta_{m}^{+(h=0)},\,\theta_{\pi}^{(h=0)}\bigr).
\)

\textbf{Perform the $s$-th EM run:}
\While{$\bigl|\,l_n\bigl(\eta^{(h+1)}\bigr)-l_n\bigl(\eta^{(h)}\bigr)\bigr| > \varepsilon$}{
\textbf{E-step:} Update $\tau_{m'}(c_r;\eta^{(h)}),\,\tau_g(c_r;\eta^{(h)}),\,\tau_m(c_r;\eta^{(h)})$ for all $r=1,\dots,n$.

\textbf{M1-step $(\mu_g, \sigma_g)$:} Maximize
\[
Q_{\phi}\bigl(\theta_{\phi},\,\eta^{(h)}\bigr)
    \;=\;
    \sum_{r=1}^{n}\sum_{g=1}^{G}\tau_g\bigl(c_r;\eta^{(h)}\bigr)\,\log\phi\bigl(c_r;\mu_g,\sigma_g\bigr)
\]
subject to
\(
\sigma_g\ge \exp\bigl(-n^d\bigr)
\)
for $g=1,\dots,G.$

\textbf{M2-step ($\pi_{m'}, \pi_g, \pi_m$):} Maximize

\[
Q_{\pi}\bigl(\theta_{\pi},\,\eta^{(h)}\bigr) = \sum_{r=1}^{n} \Bigg[ \sum_{m'=1}^{M^-} \tau_{m'}(c_r; \eta^{(h)}) \log \pi^-_{m'} 
+ \sum_{g=1}^{G} \tau_{g}(c_r; \eta^{(h)}) \log \pi_g + \sum_{m=1}^{M^+} \tau_{m}(c_r; \eta^{(h)}) \log \pi_m^+ \Bigg]
\]
subject to $\pi^-_{1} \geq \pi^-_{2} \geq \dots \geq \pi^-_{M^-}$ and $\pi^+_{1} \geq \pi^+_{2} \geq \dots \geq \pi^+_{M^+}$.}

Set
\[
\eta^{(s)} \;\gets\;\eta^{(h+1)}.
\]
}

\KwOut{$\eta^{(s^*)}$ \text{ such that } $s^* = \arg\max\limits_{s} l_n\bigl(\eta^{(s)}\bigr)$}
\end{algorithm}

\clearpage

\section{Frequency Model}
\label{sec:frequency-model}

\par This section proposes trip- and driver-level risk indices by combining portfolio-level severity layers from Section~\ref{sec:severity-model} with trip-level layer frequencies to be modeled in this section. Because the severity is estimated at the portfolio level, all trips share the same layers and corresponding severity weights, which makes the resulting indices comparable across trips and drivers. For each trip, we propose to use Poisson-Gamma conjugacy to model how often its aggregated coefficients fall into each severity layer and use this frequency profile to form a joint frequency–severity risk rate. The model admits sequential Bayesian updating as new trips arrive, which yields a driver-level index that evolves over time.

\subsection{Trip-Level Risk Index}
\par In Section~\ref{sec:dwt}, we introduced the series of aggregated wavelet coefficients  for trip $i$ as $\mathbf{C}_i=\{C_{it}\}^{T_i-1}_{t=0}$, where $t=0,\ldots,T_i-1$ indexes time points and $\mathcal{T}_i$ is the set of retained coefficients obtained by the proposed sampling scheme. In this section, we set $E_i:=|\mathcal{T}_i|$ to interpret as an \emph{exposure} measure. Meanwhile, Section~\ref{sec:severity-model} yields a collection of $M^-+M^+$ disjoint Uniform components that define severity layers, with supports $\{\theta^-_{m'}\}_{m'=1}^{M^-}$ on the left tail and $\{\theta^+_{m}\}_{m=1}^{M^+}$ on the right tail. For notational convenience, we combine and re-index these tail layers as $\{\Theta_m\}_{m=1}^{M}$, where $M=M^-+M^+$, by
\[
\Theta_m =
\begin{cases}
\theta^-_{m}=(u^-_{m+1},u^-_{m}), & m=1,\ldots,M^-,\\[4pt]
\theta^+_{m-M^-}=(u^+_{m-M^-},u^+_{m-M^-+1}), & m=M^-+1,\ldots,M.
\end{cases}
\]
For each individual trip $i$, we want to examine the frequency profile for each severity layer, so we define the trip-level \emph{multi-layer tail counts (MLTC)} by
\begin{equation}
\label{eq:mltc}
    N_{im} \;:=\; \sum_{t=0}^{E_i-1}\mathbf{1}\{\,C_{it}\in\Theta_m\,\}, \qquad m=1,\ldots,M,
\end{equation}
where $N_{im}$ counts the number of coefficients that fall in layer $\Theta_m$. The vector $\mathbf{N}_i=(N_{i1},\ldots,N_{iM})$ summarizes the tail profile across severity layers for the trip $i$. Because the layers ${\Theta_m}$ are disjoint, $\mathbf{N}_i$ captures not only how often tail patterns occur, but also how the tail counts are distributed across depths.

\par A Poisson model with exposure offset $E_i$ provides a simple way to model MLTCs and to compare trips of different lengths through per-exposure intensities $\lambda_{im}$. Conditional on the per-exposure intensities $\lambda_{im}>0$, we therefore propose to model the MLTC by
\[
N_{im}\mid \lambda_{im},E_i \;\sim\; \mathrm{Poisson}(E_i\,\lambda_{im}),
\qquad m=1,\ldots,M.
\]
To update the risk profile using a closed-form, we propose to use independent Gamma priors on intensities $\lambda_{im}$ for each severity layer $m$ as
\[
\lambda_{im}\sim \mathrm{Gamma}(\alpha_{0m},\beta_{0m}),
\qquad m=1,\ldots,M,
\]
which yields conjugate posteriors and supports stable trip- and sequential driver-level risk evaluation. By Gamma--Poisson conjugacy, the intensity is updated with the information $(E_i, \mathbf{N}_i)$ of trip $i$:
\[
\lambda_{im}\mid E_i,N_{im}
\;\sim\;
\mathrm{Gamma}(\alpha_{0m}+N_{im},\;\beta_{0m}+E_i),
\qquad m=1,\ldots,M.
\]
The layer-wise hyperparameters $(\alpha_{0m},\beta_{0m})$ are initialized by a
\emph{winsorized-moment empirical Bayes} step \citet{phipson2016robust} to reduce the influence of extreme rates on the prior fit. For each layer $m$, we compute per-exposure rates $N_{im}/E_i$ across trips and replace the values outside the $[\Omega, 1-\Omega]$ empirical quantile range with the quantile cutoffs $\Omega \in \mathbb{R}$. We then initialize the hyperparameters $(\alpha_{0m}, \beta_{0m})$ by moment matching using the sample mean and variance of the winsorized rates.

\par To penalize severity layers based on their portfolio-level rarity, we define \emph{severity weights} by
\begin{equation}
w_m
\;:=\;
\frac{(1/\pi_m)^{\gamma}}{\sum_{l=1}^M (1/\pi_l)^{\gamma}},
\qquad m=1,\ldots,M,
\label{eq:severity}
\end{equation}
where the membership probabilities $\{\pi_m\}_{m=1}^M$ are estimated by the severity model and $\gamma\ge 0$ is fixed. The parameter $\gamma$ controls the dispersion of severity weights over the layers. Smaller $\gamma$ spreads severity weights more evenly across layers, while larger $\gamma$ puts more weight on deeper layers.

\par For the trip $i$, we now propose the \emph{trip-level risk index} that summarizes severity-weighted layer-wise risk rates per exposure:
\[
S_i \;:=\; \sum_{m=1}^M w_m\,\lambda_{im}.
\]
This index aggregates the per-exposure intensities $\{\lambda_{im}\}_{m=1}^M$ of an individual trip, while using the portfolio-level severity weights $\{w_m\}_{m=1}^M$ that commonly penalize the layers across trips. Since we assume conditional independence of $N_{im}$ given $(\lambda_{im}, E_i)$ and independent Gamma priors across layers $m$, we can estimate $S_i$ by a sum of layer-wise posterior means:
\begin{equation}
\label{eq:trip_risk_rate}
\widehat{S}_i
\;=\;
\mathbb{E}[S_i\mid E_i,\mathbf{N}_i]
\;=\;
\sum_{m=1}^M w_m\,
      \frac{\alpha_{0m} + N_{im}}{\beta_{0m} + E_i},
\end{equation}
where $\dfrac{\alpha_{0m}+N_{im}}{\beta_{0m}+E_i} = \mathbb{E}[\lambda_{im}\mid E_i,\mathbf{N}_i]$. 
The estimated index $\widehat{S}_i$ admits the layer-wise decomposition so each term $w_m\,\mathbb{E}[\lambda_{im}\mid E_i,\mathbf{N}_i]$ quantifies the contribution of severity layer $m$ to the overall risk. Consequently, $\widehat{S}_i$ jointly combines the frequency information in the MLTC profile $\mathbf{N}_i$ with the portfolio-level severity weights $\{w_m\}^M_{m=1}$ on a common per-exposure scale through $E_i$, yielding a frequency--severity joint trip-level risk index.

\subsection{Driver-Level Risk Index}
\par Once the trip-level index is defined, we extend this to driver-level through sequential updating of $\lambda_{im}$ over a driver’s trips. Suppose trips of driver $d$ are observed in time order $i=1,\ldots,k$, and trip $i$ provides exposure $E_i$ and MLTC $\mathbf{N}_{i} = (N_{i1}, \dots, N_{iM})$. For each layer $m$, let $\lambda_{dm}$ denote the driver’s per-exposure intensity of tail layer $m$. After $k$ trips, conjugacy gives
\[
\lambda_{dm} \mid \mathcal{H}_d^{(k)} \sim \mathrm{Gamma}(\alpha_{dm}^{(k)},\beta_{dm}^{(k)}), \qquad \mathcal{H}_d^{(k)}=\{(E_i,\mathbf{N}_i): i=1,\ldots,k\},
\]
with sequential updates
\[
\alpha_{dm}^{(k)}=\alpha_{dm}^{(k-1)}+N_{km}, \qquad \beta_{dm}^{(k)}=\beta_{dm}^{(k-1)}+E_k,
\]
where $\alpha_{dm}^{(0)}=\alpha_{0m},\ \beta_{dm}^{(0)}=\beta_{0m}$ initialized in the same way for the trip-level index.

\par To estimate a driver’s long-term risk profile, we summarize each layer intensity $\lambda_{dm}$ after $k$ trips by its posterior mean. Under the Poisson--Gamma model, this model choice yields a closed-form update that can be implemented online as each new trip arrives. Moreover, the posterior mean admits a credibility representation because it blends the accumulated history $\widehat{\lambda}^{(k-1)}_{dm}$ with the new trip’s empirical rate $N_{km}/E_k$ through exposure-based weights. The posterior mean intensity after $k$ trips is
\begin{equation}
\label{eq:posterior_intensity}
\widehat{\lambda}_{dm}^{(k)} \;:=\; \mathbb{E}\!\left[\lambda_{dm}\mid \mathcal{H}_d^{(k)}\right] \;=\; \frac{\alpha_{dm}^{(k)}}{\beta_{dm}^{(k)}} \;=\; \underbrace{\frac{\beta_{dm}^{(k-1)}}{\beta_{dm}^{(k-1)}+E_k}}_{\text{weight on past}}
\,\widehat{\lambda}_{dm}^{(k-1)} \;+\; \underbrace{\frac{E_k}{\beta_{dm}^{(k-1)}+E_k}}_{\text{weight on new}} \,\frac{N_{km}}{E_k},
\end{equation}
where the weights depend on accumulated precision and exposure. In particular, larger $E_k$ increases the influence of the new trip, whereas small $E_k$ yields a more conservative update driven by the past through $\beta_{dm}^{(k-1)}$.

\par Given the updated layer intensities, we estimate the \emph{driver-level risk index} after $k$ trips:
\begin{equation}
\widehat{S}_d^{(k)} \;:=\; \sum_{m=1}^M w_m\,\widehat{\lambda}_{dm}^{(k)}
= \sum_{m=1}^M w_m\,\frac{\alpha_{dm}^{(k)}} {\beta_{dm}^{(k)}}.
\label{eq:cum_risk_rate}
\end{equation}
This index provides a time-evolving summary of the driver’s overall risk profile, aggregating evidence accumulated across trips. The trip-level index is recovered as the special case $k=1$, then $\alpha^{(1)}_{dm} = \alpha_{0m} + N_{1m}$ and $\beta^{(1)}_{dm} = \beta_{0m} + E_{1}$, so $\hat{S}^{(1)}_d$ matches \eqref{eq:trip_risk_rate}.

\par In Section~\ref{sec:application}, we will use these outputs: MLTC profiles $\mathbf{N}_i$, trip-level indices $\widehat{S}_i$, and sequentially updated driver-level indices $\widehat{S}^{(k)}_d$, to examine how different driving states exhibit distinct profile in terms of both frequency and severity and to evaluate how well the proposed risk index distinguishes risky from normal trips in classification.

\section{Application to Controlled Experiments}
\label{sec:application}

\par This section applies the proposed frequency--severity risk index framework to the UAH-DriveSet controlled dataset described in Section~\ref{sec:intro}. We begin by selecting the MODWT decomposition depth $J$, constructing a balanced portfolio sample of aggregated wavelet coefficients by using the proposed sampling scheme discussed in Section~\ref{sec:severity-model}, and specifying the model components and tuning grids needed for estimation. We then report the resulting fitted model (including the $\gamma$ selected and the corresponding severity weights) and interpret it both visually and numerically, and illustrate how the proposed representation captures abnormal driving patterns across diverse time scales. Based on the fitted results, we examine MLTC profiles for each trip together with the associated trip- and sequentially updated driver-level indices. Finally, we evaluate how well the proposed index improves discrimination between normal and risky trips in a binary classification analysis.

\subsection{Portfolio Construction and Model Setup}
\par
\begin{figure}[H]
  \centering  \includegraphics[height=0.60\textheight,width=0.75\textwidth]{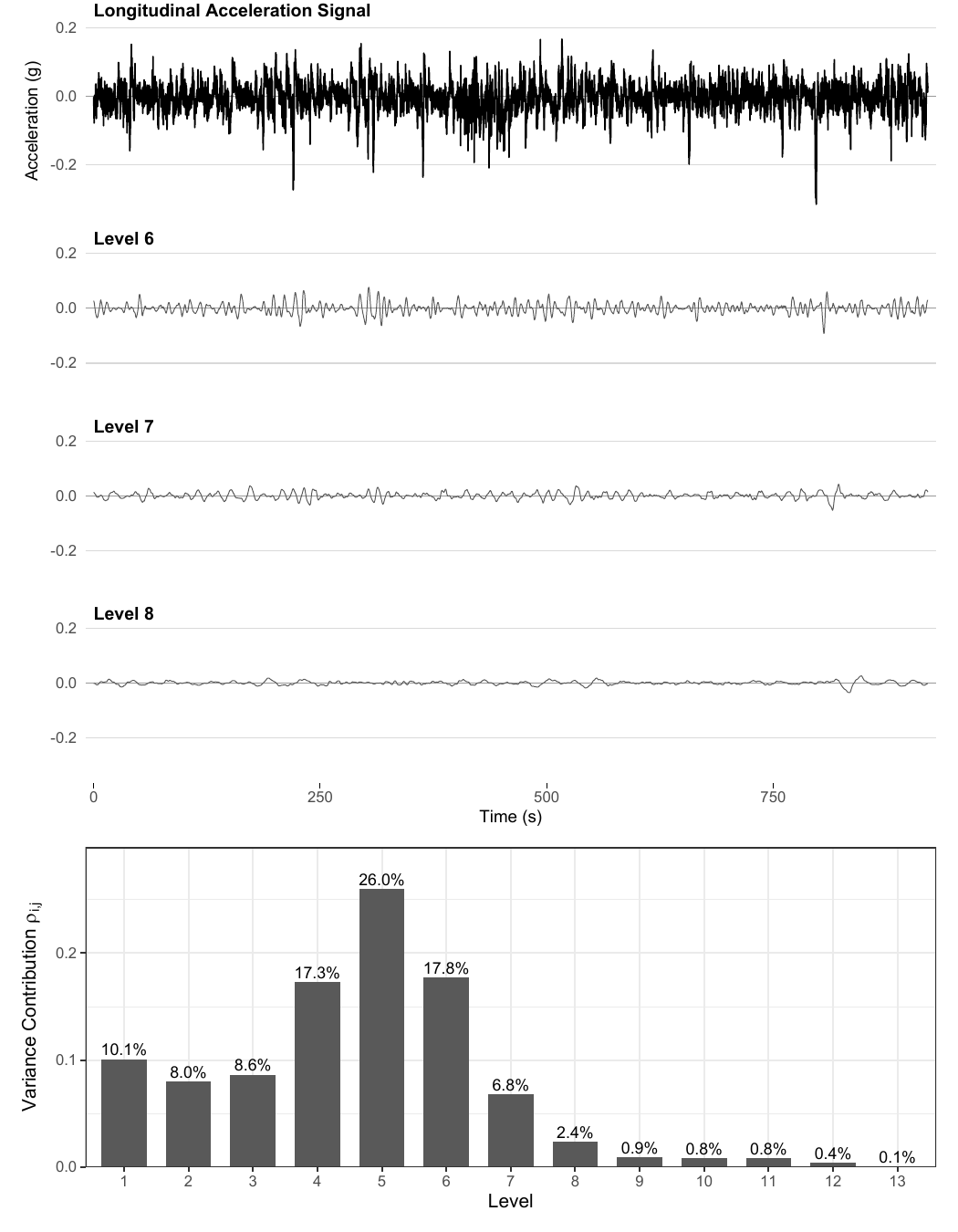}
  \caption{Selecting the decomposition depth $J$: The raw longitudinal acceleration for an \texttt{aggressive} trip of Driver 4 on Motorway (top), level 6/7/8 MODWT coefficient trajectories (middle), and the corresponding variance contributions across levels (bottom), illustrating that risk-relevant variations become sparse at deeper levels and motivating the choice of $J{=}6$.}
  \label{fig:justification-J}
\end{figure}

 We select the MODWT decomposition depth $J$ using the variance contribution ${\rho_{i,j}}$ defined in Section~\ref{sec:dwt} and a visual inspection of the MODWT coefficient trajectories across levels so that we retain as much risk-relevant information as possible while keeping the representation computationally efficient. For this, we begin from a conservative reference trip by Driver 4, the \texttt{aggressive} trip on motorway, because it has the longest trip length and therefore admits the largest possible decomposition depth under the conventional rule for a full depth, $J_{\text{full}}=\lfloor \log_2(T_i)\rfloor$. For this trip, the cumulative variance reaches about $99.2\%$ at $J_{\text{full}}=13$, while truncation at $J=6$ and $J=7$ retains about $87.8\%\ (=\sum^6_{j=1}\rho_{i,j})$ and $94.6\%\ (=\sum^7_{j=1}\rho_{i,j})$ of the sample variance $\mathbf{X}_i$, respectively. Figure~\ref{fig:justification-J} shows a sharp drop in the incremental variance contribution when moving from level $6$ to level $7$, and the coefficient trajectories likewise indicate that, from level $7$, pronounced risk-relevant variations become significantly rare compared with the lower scales. We therefore select $J=6$ as a conservative and efficient choice.

\par Following the sampling scheme proposed in Section~\ref{sec:severity-model}, we construct the portfolio sample $\mathbf{C}$ from the aggregated coefficient series $\mathbf{C}_i=\{C_{it}\}_{t=0}^{T_i-1}$. In our controlled data setting, the driving tests are conducted under well-controlled experimental conditions, such as on pre-specified routes and in traffic/signal environments that allow continuous and natural driving, so the behavior-irrelevant noises discussed in Section~\ref{sec:severity-model} are largely absent. Accordingly, at the hierarchical sampling stage, we decide to use all $40$ available trips. On the other hand, at the thinning stage, we aim to reduce the strong serial dependence of $\{C_{it}\}$ so that the pooled sample is closer to an approximately i.i.d.\ input for likelihood fitting. For each trip $i$, we compute the sample autocorrelation function (ACF) of $\mathbf{C}_i$ and choose the thinning lag $\texttt{lag}_i$ as the smallest lag such that the absolute ACF stays below $0.1$ for three consecutive lags (at $\texttt{lag}_i,\texttt{lag}_i+1,\texttt{lag}_i+2$). Given $\texttt{lag}_i$, we retain coefficients at time point
$t_i,\, t_i+\texttt{lag}_i,\, t_i+2\texttt{lag}_i,\ldots$,
where the start $t_i\sim \mathrm{Uniform}\{0,1,\ldots,\texttt{lag}_i-1\}$ is randomized to avoid a fixed sampling phase across trips. Pooling the retained coefficients across trips yields the portfolio sample
$\mathbf{C}=\bigcup_i\{C_{it}: t\in\mathcal{T}_i\}$. Across the $40$ trips, the selected lags, on average, about $8$ time points (with $0.1$-second resolution), and thinning reduces the pooled sample size from $311{,}381$ to $38{,}219$. After thinning, we define the exposure for trip $i$ as the number of retained coefficients, $E_i:=|\mathcal{T}_i|$, so the subsequent MLTC profiles and risk indices are computed on the same thinned support used to construct the portfolio distribution.

\par We now fit our severity model to the resulting portfolio sample $\mathbf{C}$ constructed based on the chosen depth $J=6$. We use the \emph{maximum choice} rule for across-level aggregation because, at each time point, it keeps the strongest coefficient observed across levels. This prevents strong patterns from being averaged out by milder ones at other levels, and thus retains the most salient risk-relevant patterns for subsequent modeling. For the model tuning parameters, we set \((d,\alpha,\delta,q,p)=(0.5,\,0.05,\,1.96,\,12,\,10)\) and fit over \(G \in \{1,2\}\), \(M^- \in \{1, 2, \dots,8\}\), and \(M^+ \in \{1, 2, \dots, 6\}\). We set \(d=0.5\) as recommended by \citet{Tanaka2006} and $\alpha=0.05$ to use the outer $5\%$ of the portfolio to form the base grid for candidate search for Uniform parameters. The grid $(q,p)$ controls the tail partition resolution for the base grid, and $(M^-, M^+)$ controls the number of severity layers on each tail. We use finer resolution on the left tail by setting $q=12$ and allowing more layers $M^- \in \{1, 2, \dots,8\}$, reflecting that the portfolio sample has a longer left tail, while we set $p=10$ with $M^+ \in \{1, 2, \dots,6\}$ on the right tail.

\subsection{Fitted Severity Model and Portfolio-Level Severity Weights}
\par Under the model setup described above, we fit the proposed MU-MEMR severity model over $(M^-,G,M^+)$. Table~\ref{tab:model_select_summary} summarizes the selected specification under three criteria, including the log-likelihood, AIC, and BIC. In what follows, we report results under $(M^-,G,M^+)=(4,2,5)$ selected by BIC. This choice avoids creating extremely sparse deep layers that would absorb disproportionately large severity weights, so the performance of the resulting risk index does not excessively rely on a few extreme layers. Using the estimated membership probabilities of the fitted layers, we then select the tuning parameter $\gamma$ which manages the dispersion of severity weigths over the layers based on the procedure in Appendix~\ref{app:classification}. With the selected $\gamma=1.7$, we compute the resulting severity weights $\mathbf{w}=(w_1,\ldots,w_M)$ and apply these weights to all trips as a fair penalty scale across trips. This makes the proposed risk index comparable across trips as a single, interpretable measure of telematics risk.

\begin{table}[H]
\centering
\caption{Severity model selection over $(M^-,G,M^+)$: the best specification under log-likelihood, AIC, and BIC for the portfolio distribution.}
\label{tab:model_select_summary}

\small
\setlength{\tabcolsep}{3pt}
\renewcommand{\arraystretch}{1.05}
\newcommand{\sr}[1]{#1}

\begin{tabularx}{0.58\linewidth}{@{}>{\raggedright\arraybackslash}X c c c r@{}}
\toprule
Criterion & $G$ & $M^-$ & $M^+$ & Value \\
\midrule
log-likelihood & \multirow{3}{*}{\sr{2}} & \sr{8} & \sr{6} & \sr{95552.84} \\
AIC        &                         & \sr{7} & \sr{6} & \sr{-191040.1} \\
BIC        &                         & \sr{4} & \sr{5} & \sr{-190816.8} \\
\bottomrule
\end{tabularx}
\end{table}

\par
\begin{figure}[H]
  \centering
  \includegraphics[width=0.9\textwidth, height=7.0cm]{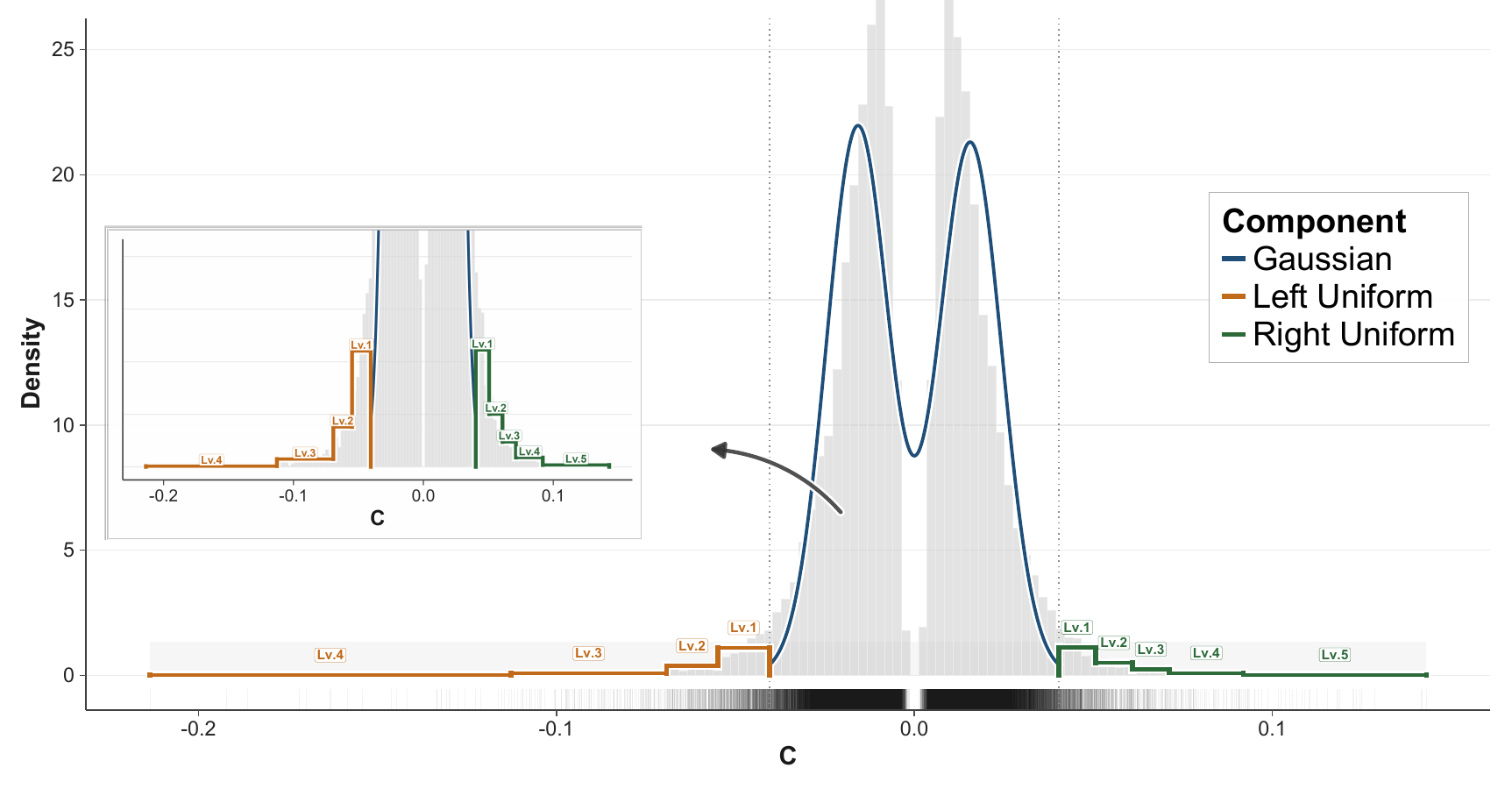}
  \caption{
  Fitted Gaussian--Uniform mixture under BIC for the portfolio distribution of aggregated wavelet coefficients: the Gaussian bulk captures typical driving dynamics, while multi-layer Uniform tails define ordered severity layers; rug marks along the axis show the assignment of individual coefficients.}
  \label{fig:fitted_mixture}
\end{figure}

 The resulting fitted severity model, shown in Figure~\ref{fig:fitted_mixture} and Table~\ref{tab:estimated_parms_severity_model}, illustrates the representativeness of the proposed Gaussian--Uniform mixture severity model as discussed in Section~\ref{sec:severity-model} and the robustness of the risk index computed by using the fitted model. Although the empirical distribution exhibits a longer left tail shown in Figure~\ref{fig:fitted_mixture}, two Gaussian components remain nearly symmetric around zero with similar variances shown in Table~\ref{tab:estimated_parms_severity_model}. This  indicates that tail outliers are not inflating the left Gaussian's scale because the Uniform absorbs the contaminated observations. This matters for risk assessment because without Uniform absorption the fitted model can treat part of the long-tail region as if it were ordinary, leading to a baseline that is too generous for tail behaviors. Given the bulk--tail separation in Figure~\ref{fig:fitted_mixture} and the supporting estimates in Table~\ref{tab:estimated_parms_severity_model}, the severity model attains the representativeness: the portfolio baseline is stably determined by typical driving dynamics, while rare deviations are isolated into tail layers. This separation provides a reliable baseline to regard the patterns in Gaussians as normal and in Uniforms as abnormal at the portfolio level. Therefore, the proposed risk index which evaluates risk through the frequency and depth of Uniform layer assignments can be viewed as capturing the genuine risk-relevant information.

\begin{table}[H]
\centering
\small

\setlength{\tabcolsep}{5pt}
\renewcommand{\arraystretch}{0.9}

\newcommand{\blockhdr}[1]{\rowcolor{black!4}\multicolumn{4}{@{}l@{}}{\textbf{#1}}\\}

\begin{tabular*}{0.70\textwidth}{@{\extracolsep{\fill}} l
S[table-format=-1.5]
S[table-format= 1.5]
@{}}
\toprule
\multicolumn{3}{@{}l@{}}{\textbf{Gaussian component} $(G=2)$} \\
\midrule
 & {$g=1$} & {$g=2$} \\
\cmidrule(lr){2-2}\cmidrule(lr){3-3}
$\mu_g$    & -0.0158  & 0.0156  \\
$\sigma_g$ & 0.00877  & 0.00883 \\

$\pi_g$    & \multicolumn{1}{c}{48.19\%} & \multicolumn{1}{c}{47.10\%} \\
\bottomrule
\end{tabular*}

\vspace{4pt}
\begin{tabular*}{0.82\textwidth}{@{\extracolsep{\fill}} l
c
c
S[table-format=1.4]
@{}}
\toprule
\multicolumn{4}{@{}l@{}}{\textbf{Uniform component} ($M^- = 4,\; M^+ = 5$)} \\
\midrule
{} & {Interval} & {$\pi_m$ (\%)} & {$w_m$} \\
\midrule

\blockhdr{Left tail layers}
$m'=4$ & $\left[\num{-0.2136},\,\num{-0.1127}\right]$ & 0.071\% & 0.4726 \\
$m'=3$ & $\left[\num{-0.1127},\,\num{-0.0693}\right]$ & 0.322\% & 0.0360 \\
$m'=2$ & $\left[\num{-0.0693},\,\num{-0.0549}\right]$ & 0.547\% & 0.0147 \\
$m'=1$ & $\left[\num{-0.0549},\,\num{-0.0405}\right]$ & 1.591\% & 0.0024 \\
\addlinespace[2pt]

\blockhdr{Right tail layers}
$m=1$ & $\left[\num{0.0403},\,\num{0.0506}\right]$   & 1.142\% & 0.0042 \\
$m=2$ & $\left[\num{0.0506},\,\num{0.0609}\right]$   & 0.516\% & 0.0162 \\
$m=3$ & $\left[\num{0.0609},\,\num{0.0713}\right]$   & 0.246\% & 0.0569 \\
$m=4$ & $\left[\num{0.0713},\,\num{0.0919}\right]$   & 0.183\% & 0.0938 \\
$m=5$ & $\left[\num{0.0919},\,\num{0.1430}\right]$   & 0.092\% & 0.3032 \\
\bottomrule
\end{tabular*}

\caption{Estimated parameters of the Gaussian--Uniform severity model based on BIC: Gaussian bulk components with their means, scales, and membership probabilities and Uniform tail layers with their intervals, layer probabilities, and severity weights.}
\label{tab:estimated_parms_severity_model}
\end{table}

\subsection{Trip- and Driver-Level Risk Indices}
\par We estimate the proposed trip- and driver-level risk indices by combining the Poisson--Gamma frequency model in Section~\ref{sec:frequency-model} with the portfolio-level severity layers and severity weights from the fitted Gaussian--Uniform mixture model in Section~\ref{sec:severity-model}. The frequency and severity models are fitted using the full set of $40$ trips pooled across both secondary and motorway roads, so the estimated portfolio baseline and severity scale reflect the entire dataset. For exposition and interpretability, we focus on the secondary road when presenting MLTC profiles and risk indices. This focus can isolate the behavioral differences and enable the characterization of the resulting frequency--severity profiles across the three driving states by examining the trips on the same route and speed regime. Moreover, D1--D5 have four trips on secondary road for each (versus three motorway trips), which provides a richer sequence within drivers and allows the sequential updating mechanism to be illustrated more clearly: normal trips tend to moderate the driver-level index, while risky trips tend to increase it.

\par From the estimated MLTC profiles and risk indices, we demonstrate both (i) how tail \emph{depth} changes ranking of trip-level indices and (ii) how sequential updating aggregates evidence into a driver-level risk profile over time. Table~\ref{tab:secondary_index_MLTC} illustrates how the multi-layer tail structure shapes risk evaluation by accounting for \emph{which} severity layers are reached, not only how many tail counts occur in total. Across drivers, the \texttt{normal} trips are characterized by either small MLTC totals and/or tail counts concentrated in shallow layers (typically Lv1$\pm$ or Lv2$\pm$), whereas the risky trips tend to exhibit either higher overall MLTC and/or a broader spread of tail mass into deeper layers. Moreover, while both risky states often extend beyond the shallow layers, \texttt{aggressive} trips typically distribute MLTC mass more heavily across multiple layers, whereas \texttt{drowsy} trips more often display a lighter overall mass but still reach deep layers.

\par Driver~6 provides an important observation: unlike Drivers~1--5, there are only two trips available and Driver~6's \texttt{normal} trip already exhibits non-negligible tail mass reaching moderately deep layers (e.g., Lv3- and Lv4+). Nevertheless, the proposed index still flags \texttt{drowsy} as riskier because the evaluation is \emph{exposure-normalized} and depth-weighted: despite having fewer total tail counts, the \texttt{drowsy} trip both occurs over a shorter exposure window and reaches the deepest left-tail layer (Lv4-), so its per-exposure contribution in the deepest layer is amplified. This illustrates that, once the portfolio severity scale is stable, the trip-level assessment can remain informative even for a new policyholder with limited trip history, because it compares trips on a common per-exposure scale while appropriately prioritizing extreme behavior.

\par This layer-wise structure characterizes interpretable frequency--severity profiles across different behavioral driving states. Here, a trip's \emph{frequency profile} refers to the overall MLTC mass, while its \emph{severity profile} describes how that tail mass is distributed across layers. Normal trips tend to be low-frequency/low-severity because most of their MLTC mass is concentrated in the shallow layers. Aggressive trips tend to be high-frequency/high-severity because they show many tail counts that spread across layers and commonly reach deeper tail layers (e.g., Lv3$\pm$ or beyond). Drowsy trips often exhibit a distinct low-frequency/high-severity profile because their counts can be similar in shallow layers compared with normal, while the observed counts are more likely to appear in deeper layers. This is exactly where the proposed framework is most informative, because drowsy driving can look ``quiet'' under the consideration of frequency only, while its high-severity profile reveals itself in the tails.

\begin{table}[H]
\centering
\renewcommand{\arraystretch}{0.92}
\setlength{\tabcolsep}{3pt}
\setlength{\extrarowheight}{-1pt}
\scriptsize

\newcommand{\zb}{} 

\begin{adjustbox}{max width=\textwidth, max totalheight=1.0\textheight, keepaspectratio}

\begin{tabular}{l l >{\color{black}}r >{\color{black}}r >{\color{black}}r *{9}{>{\color{black}}r}}
\toprule
\multirow{2}{*}{\textbf{Driver}} & \multirow{2}{*}{\textbf{Label}} &
\multirow{2}{*}{\textbf{Exposure} ($E_i$)} &
\multicolumn{2}{c}{\textbf{Risk Index} ($\times10^{-4}$)} &
\multicolumn{9}{c}{\textbf{MLTC} ($M^-{=}4,\,M^+{=}5$)} \\
\cmidrule(lr){4-5}\cmidrule(l){6-14}
 &  &  &
\textcolor{black}{\texttt{Trip}} & \textcolor{black}{\texttt{Updated}} &
\multicolumn{4}{c}{\textit{Left tail(-)}} & \multicolumn{5}{c}{\textit{Right tail(+)}} \\
 &  &  &  &  &
\textcolor{black}{Lv4-} & \textcolor{black}{Lv3-} & \textcolor{black}{Lv2-} & \textcolor{black}{Lv1-} &
\textcolor{black}{Lv1+} & \textcolor{black}{Lv2+} & \textcolor{black}{Lv3+} & \textcolor{black}{Lv4+} & \textcolor{black}{Lv5+} \\
\midrule

\multirow{4}{*}{\textbf{D1}}
 & normal1    & 685  & 4.05 & 4.05  & \zb & \zb & \zb & \zb & \zb & 1  & \zb & \zb & \zb \\
 & aggressive & 759  & \textbf{19.04} & 11.38 & \zb & 5  & 9  & 10 & 13 & 6  & 6  & 1  & 2  \\
 & normal2    & 725  & 5.92 & 8.81  & \zb & 1  & 4  & 6  & 5  & 1  & 1  & \zb & \zb \\
 & drowsy     & 1004 & \textbf{42.13} & 20.80 & 4  & 10 & 7  & 16 & 16 & 4  & 4  & 5  & 6  \\
\midrule

\multirow{4}{*}{\textbf{D2}}
 & normal1    & 724 & 3.93 & 3.93  & \zb & \zb & 1  & 2  & \zb & \zb & \zb & \zb & \zb \\
 & aggressive & 617 & \textbf{47.20} & 25.72 & 1  & 13 & 25 & 41 & 37 & 11 & 10 & 5  & 3  \\
 & normal2    & 763 & 4.25 & 17.63 & \zb & \zb & 1  & 3  & 6  & 1  & \zb & \zb & \zb \\
 & drowsy     & 930 & \textbf{47.14} & 28.72 & 5  & 10 & 6  & 18 & 23 & 4  & 3  & 10 & 4  \\
\midrule

\multirow{4}{*}{\textbf{D3}}
 & normal1    & 874 & 3.40 & 3.40  & \zb & \zb & \zb & 3  & 2  & \zb & \zb & \zb & \zb \\
 & aggressive & 678 & \textbf{11.28} & 5.87 & \zb & 1  & 5  & 15 & 7  & 8  & 3  & 2  & \zb \\
 & normal2    & 835 & 3.96 & 4.24  & \zb & \zb & \zb & 14 & 3  & 1  & \zb & \zb & \zb \\
 & drowsy     & 828 & \textbf{10.33} & 5.44 & 1  & 2  & 1  & 8  & 2  & \zb & \zb & 2  & \zb \\
\midrule

\multirow{4}{*}{\textbf{D4}}
 & normal1    & 810 & 3.75 & 3.75  & \zb & \zb & \zb & 3  & 2  & 1  & \zb & \zb & \zb \\
 & aggressive & 828 & \textbf{15.24} & 8.91 & 1  & 1  & 3  & 17 & 23 & 2  & 3  & 1  & 1  \\
 & normal2    & 863 & 4.99 & 6.80  & \zb & \zb & 2  & 3  & 3  & 1  & 2  & \zb & \zb \\
 & drowsy     & 859 & \textbf{11.52} & 7.78 & 1  & 1  & 4  & 2  & 10 & 2  & \zb & \zb & 1  \\
\midrule

\multirow{4}{*}{\textbf{D5}}
 & normal1    & 875 & 4.33 & 4.33  & \zb & \zb & \zb & 1  & 2  & 1  & \zb & 1  & \zb \\
 & aggressive & 283 & \textbf{58.38} & 23.90 & 4  & 3  & \zb & 5  & 4  & 1  & 4  & 1  & 2  \\
 & normal2    & 989 & 4.45 & 14.98 & \zb & \zb & 4  & 9  & 6  & 5  & \zb & \zb & \zb \\
 & drowsy     & 876 & \textbf{8.64} & 12.87 & \zb & 4  & 3  & 8  & 1  & 2  & 1  & \zb & 1  \\
\midrule

\multirow{2}{*}{\textbf{D6}}
 & normal     & 1947 & 5.96 & 5.96 & \zb & 5  & 9  & 42 & 21 & 6  & 4  & 1  & \zb \\
 & drowsy     & 900  & \textbf{9.38} & 6.55 & 1  & \zb & 4  & 14 & 15 & 3  & 1  & \zb & \zb \\
\bottomrule
\end{tabular}
\end{adjustbox}

\caption{Numerical result on secondary road: \textbf{Risk Index}: \texttt{Trip} (trip-level) and \texttt{Updated} (driver-level) with exposure given by the number of retained coefficients $|\mathcal{T}_i|$ and \textbf{MLTC} counts in the left tail (Lv4-$\sim$Lv1-) and right tail (Lv1+$\sim$Lv5+) under the layers \((M^-{=}4,\,M^+{=}5)\). Zeros in the MLTC columns are left blank.}
\label{tab:secondary_index_MLTC}
\end{table}

\par Once the frequency model estimates layer-wise risk rates $\lambda_{im}$ for a trip, we compute the trip-level risk index (\texttt{Trip}) by averaging these rates with the portfolio-level severity weights, i.e., a severity-weighted average of risk rates. Table~\ref{tab:secondary_index_MLTC} shows that this construction yields a clear and consistent separation between normal and risky trips across the entire sample: the largest trip-level index among normal trips is $5.96$ (Driver~6 \texttt{normal}), whereas the smallest index among risky trips is $8.64$ (Driver~5 \texttt{drowsy}), so every risky trip is ranked above every normal trip. This comparability can be interpreted directly from the MLTC profiles, and this holds even for trips with the same label. For instance, \texttt{drowsy} for Driver~2 (47.14) is far riskier than \texttt{drowsy} for Driver~5 (8.64): Driver~2’s \texttt{drowsy} exhibits heavier and deeper tail counts (Lv4- $=5$, Lv3- $=10$, Lv4+ $=10$, Lv5+ $=4$), whereas Driver~5’s \texttt{drowsy} has a much lighter MLTC profile with only a small presence in the deepest right-tail layer (Lv5+ $=1$). Thus, the index provides a numerical ordering that remains meaningful across trips because it is anchored to the same portfolio-level severity scale, enabling coherent ranking not only across different behavioral driving states but also \emph{within} the same state by distinguishing relatively less risky versus more risky trips. 

\par The MLTC breakdown also reveals which \emph{side} of the tail drives risk, linking back to the physical interpretation of longitudinal acceleration. For example, Driver~3 \texttt{drowsy} drives its risk from the left-tail profile, with the left-tail counts (Lv4- $=1$, Lv3- $=2$, Lv2- $=1$, Lv1- $=8$; total $12$) exceeding the right-tail counts (Lv1+ $=2$, Lv4+ $=2$; total $4$). This indicates that abnormal deceleration-type patterns contribute more to its risk than acceleration-type patterns, helping interpret which tail behavior drives the risk index of a trip beyond the single scalar index.

\par For pricing and underwriting, it is essential to assess a driver’s overall risk profile. We compute a driver-level (\texttt{Updated}) risk index via sequential updating to quantify how a driver’s overall profile evolves as trips accumulate. We illustrate this sequential updating dynamic by assuming an illustrative order of trips for each driver. Specifically, we arrange the four secondary road trips as \texttt{normal1} $\rightarrow$ \texttt{aggressive} $\rightarrow$ \texttt{normal2} $\rightarrow$ \texttt{drowsy} because this alternating normal--risky pattern can demonstrate how the driver-level profile reacts as different risk evidence is incorporated over time.

\par Under this illustrative order, Table~\ref{tab:secondary_index_MLTC} shows that the updated index typically increases when a risky trip is incorporated and is moderated when a normal trip is incorporated (e.g., Driver~1: $4.05 \rightarrow 11.38 \rightarrow 8.81 \rightarrow 20.80$; Driver~2: $3.93 \rightarrow 25.72 \rightarrow 17.63 \rightarrow 28.72$). To understand the updating mechanism inside at the layer level, consider Driver~5 around the transition \texttt{normal2}$\rightarrow$\texttt{drowsy}. Even though \texttt{drowsy} is a risky trip, the driver-level index can decrease if the newly observed trip implies \emph{milder} intensities than its history: before incorporating \texttt{drowsy}, the baseline of posterior intensity for Driver~5 is overly inflated in some deep layers due to the earlier \texttt{aggressive} trip, but the subsequent \texttt{drowsy} trip is not as extreme in those deep layers and therefore does not reinforce the inflated baseline. For example, when we denote $k=3$ and $k=4$ by $\{\texttt{normal1}, \texttt{aggressive}, \texttt{normal2}\}$ and $\{\texttt{normal1}, \texttt{aggressive}, \texttt{normal2}, \texttt{drowsy}\}$, respectively, the posterior intensities decrease from $k=3$ to $k=4$ in Lv4-,
$\widehat{\lambda}^{(k=3)}_{D5,\mathrm{Lv4-}}=17.41 \rightarrow \widehat{\lambda}^{(k=4)}_{D5,\mathrm{Lv4-}}=2.60$, as well as in Lv3+, $\widehat{\lambda}^{(k=3)}_{D5,\mathrm{Lv3+}}=19.18 \rightarrow \widehat{\lambda}^{(k=4)}_{D5,\mathrm{Lv3+}}=13.73$, and Lv4+,
$\widehat{\lambda}^{(k=3)}_{D5,\mathrm{Lv4+}}=9.95 \rightarrow \widehat{\lambda}^{(k=4)}_{D5,\mathrm{Lv4+}}=3.41$, in units of $10^{-4}$. Although other layers increase, the severity-weighted aggregation can still move downward because the layers that decrease from $k=3$ to $k=4$ are those carrying the majority of the severity weights (e.g., Lv4- and the upper-tail layers Lv3+/Lv4+), so their reductions dominate the smaller increases in more lightly weighted layers. This relative mildness of the deep layers means that the new trip provides insufficient evidence to reach to the previously elevated intensities in deep layers, so the posterior means in those layers revert downward and the weighted index can decrease.

\subsection{Binary Classification Validation: \texttt{normal} vs \texttt{risky}}
\par In this subsection, we demonstrate how effective the proposed frequency--severity joint framework is at identifying risky trips by a binary classification task (\texttt{normal} vs.\ \texttt{risky}) at the trip level; see Appendix~\ref{app:classification} for algorithmic details. We use all $40$ trips from the controlled dataset, and compare three MLTC-based inputs under the same classifier: Model~(A) \textbf{Total frequency (single layer)} uses the scalar total MLTCs $N_{i1}=\sum_{m=1}^M N_{im}$ (i.e., $M=1$ with weight $w_1\equiv 1$), capturing total frequency of abnormal patterns only. Model~(B) \textbf{Layer-wise frequency (unweighted)} uses the full MLTC vector $\mathbf{N}_i=(N_{i1},\ldots,N_{iM})$ with uniform weights $w_m\equiv 1/M$, preserving the multi-layer structure without severity weighting. Model~(C) \textbf{Severity-weighted frequency (proposed)} uses the same MLTC vector but sets $w_m=w_m(\gamma^\star)$ from the fitted severity model, with $\gamma^\star=1.7$ chosen by an evaluation scheme to maximize balanced accuracy; see Appendix~\ref{app:classification}. The step (A)$\to$(B) isolates the gain from retaining the layer structure beyond a single total count, and the step (B)$\to$(C) isolates the gain from incorporating portfolio-level severity weights into risk evaluation.

\begin{figure}[H]
  \centering
  \includegraphics[width=0.90\textwidth]{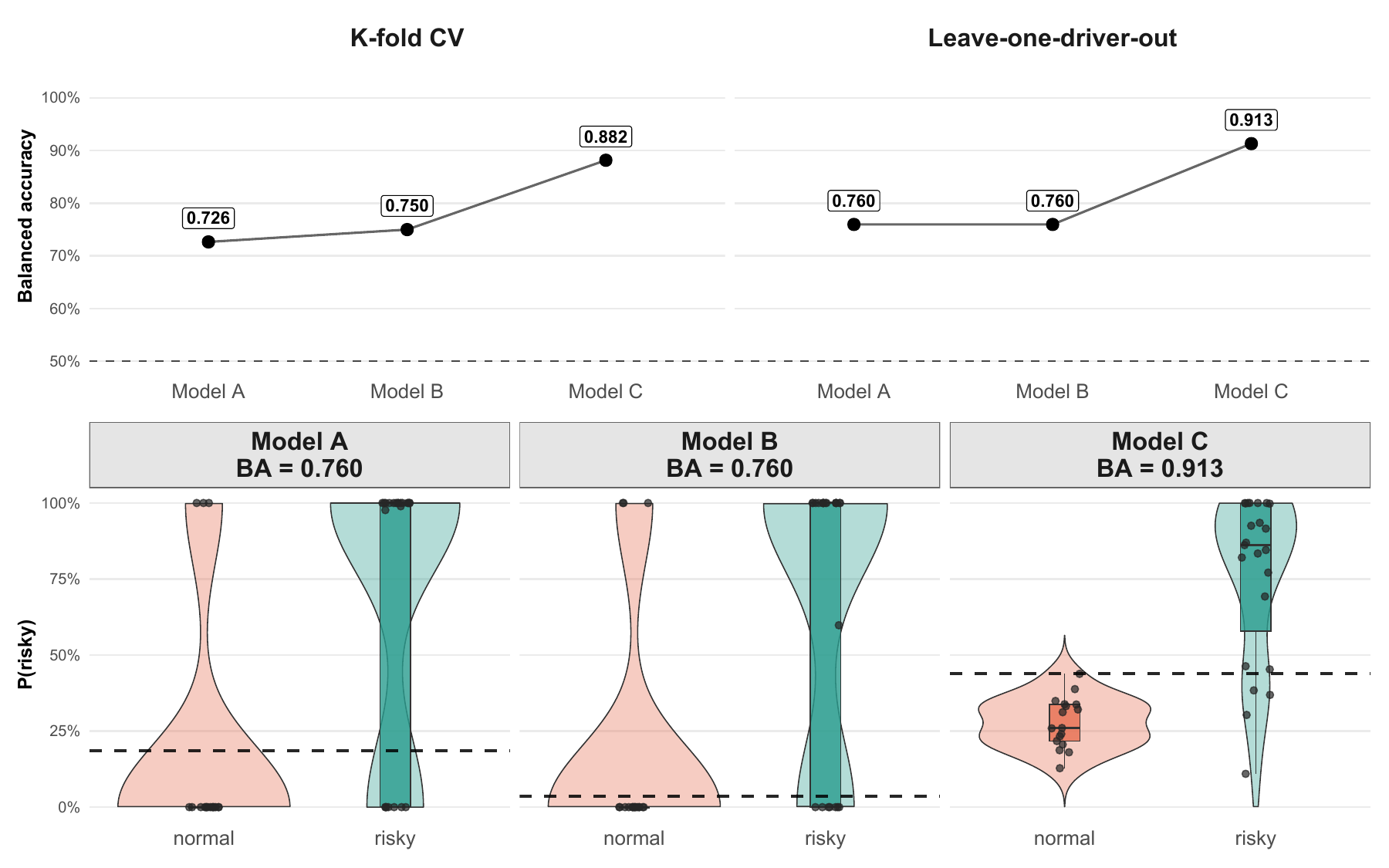}
  \caption{Binary classification (\texttt{normal} vs.\ \texttt{risky}) using MLTC-based inputs: balanced accuracy under repeated stratified $K$-fold CV and LODO (top), and out-of-fold distributions of $P(\texttt{risky})$ with the nested-CV threshold (bottom), comparing \textbf{Total frequency}, \textbf{Layer-wise frequency}, and \textbf{Severity-weighted frequency} representations.
}
  \label{fig:classification}
\end{figure}

\par Figure~\ref{fig:classification} summarizes performance in terms of balanced accuracy (BA), with two evaluation schemes. Under repeated stratified $K$-fold cross validation ($K$-fold CV), the mean BA improves from Model~A to~B to~C,
$72.6\% \rightarrow 75.0\% \rightarrow 88.2\%$. To further guard against driver leakage across trips from the same driver and to assess the framework to an unseen driver, we also report leave-one-driver-out cross validation (LODO CV), where the test fold contains all trips from one held-out driver. Under LODO, Models~A and~B both achieve BA $76.0\%$, whereas Model~C achieves BA $91.3\%$ in this controlled dataset. This monotone pattern, which holds under both repeated stratified $K$-fold CV and the stricter driver-held-out LODO setting, supports the two intended gains of our ablation design. The ablations show that preserving the multi-layer structure and then severity weighting are both crucial for improving discrimination beyond total frequency counting.

\par To visualize how well the three inputs separate \texttt{normal} and \texttt{risky} trips, the lower panel of Figure~\ref{fig:classification} compares the out-of-fold distributions of $P(\texttt{risky})$ for the total $40$ trips and highlights both the between-class separation and the within-class concentration. Model~C shows the clearest improvement: incorporating severity weights shifts \texttt{normal} trips further downward so they place at clearly lower $P(\texttt{risky})$ values than in Models~A/B, and the \texttt{normal} distribution also becomes more tightly clustered. Likewise, the \texttt{risky} distribution shifts upward under Model~C, so trips that had relatively low $P(\texttt{risky})$ under Models~A/B (and were thus prone to misclassification) move into a more decisively risky range. Importantly, the within-class spread is reduced for both classes, indicating that severity weighting stabilizes the $P(\texttt{risky})$ by reflecting not only the frequency of MLTCs but also the extremeness of the layers in which they occur. Overall, the distributional view reinforces the BA results: severity weighting is the key step that turns MLTC layer information into a more separated and more robust indicator for identifying risky trips. This sharper separation also suggests a clearer path for future work to relate the telematics-based risk index to an accident or  collision risk to uncover their relationship.

\section{Conclusion and Further Work}
\par This paper proposes a frequency--severity joint trip- and driver-level risk index for telematics signals, where severity is defined as the rarity of abnormal driving patterns relative to a portfolio-level baseline rather than as a claim size. Using a multi-scale MODWT representation of longitudinal acceleration, we learn a stable portfolio baseline and translate tail rarity into an interpretable severity scale that can be combined with trip-level tail frequencies.

\par Our main contributions are threefold. First, we formalize a portfolio-level notion of severity for telematics-based ratemaking by treating extreme tail behaviors as the risk-relevant events. Second, we extend the existing single-Uniform Gaussian--Uniform formulation to a multi-Uniform layered tail structure and develop the MU-MEMR fitting procedure to make likelihood-based estimation feasible under multiple tail layers. Third, the fitted layers induce multi-layer tail counts (MLTCs) that admit a conjugate Poisson--Gamma frequency model, yielding a closed-form per-exposure risk index and enabling sequential updating to construct dynamically evolving driver-level risk profiles.

\par In the UAH-DriveSet controlled study, the fitted severity model provides a robust separation between typical (Gaussian) dynamics and abnormal (Uniform) tail behaviors, which makes it coherent to penalize the driving patterns based on their own extremeness in risk evaluation. Empirically, MLTC layering clarifies why total tail counts alone can be misleading, while the severity-weighted index produces consistent separation between \texttt{normal} and \texttt{risky} trips and supports coherent ranking and updating at the driver level. The accompanying binary classification study further validates that incorporating portfolio-level severity weights improves discrimination, compared with other two frequency-based models (Model A and B).

\par Looking ahead, several natural extensions can further strengthen and expand the proposed framework. First, rather than treating nearby tail observations as separate MLTCs, abnormal patterns could be clustered into event-level episodes, yielding an event-based representation of risk. Such an event formulation would create a more direct bridge to downstream outcomes (e.g., collision occurrence or accident severity), enabling the study of which types of abnormal driving episodes are most predictive of real-world risk. Second, temporal dependence within a trip could be modeled explicitly through hidden Markov or other state-space formulations. Incorporating dynamic structure would allow the framework to capture regime shifts and persistence in driving behavior, rather than constructing the portfolio baseline solely through hierarchical aggregation and thinning steps. Third, the current univariate index based on longitudinal acceleration can be naturally extended to a multivariate setting. Applying the same portfolio-anchored frequency–severity construction to multiple telematics signals (e.g., speed, lateral acceleration, contextual features) and combining signal-specific indices into a unified composite measure would enable richer and more comprehensive driver risk profiling.

\par More broadly, the proposed index should be viewed not as a terminal model, but as a foundational building block for a broader class of signal-based risk assessment tools. By quantifying abnormal behavior as severity-weighted rarity at the portfolio level, the framework unlocks a range of potential applications. In insurance, it supports enriched telematics-based risk assessment, dynamic and usage-based pricing, and the construction of continuously updated driver risk profiles. It also opens the door to studying causal risk factors for hazardous driving, linking event-level behavior to accident propensity, and inferring collision severity from high-frequency telematics data.

\par Beyond insurance, aggregated trip-level scores can be used to construct spatial risk maps, identify high-risk road segments and intersections, and inform transportation safety interventions. With streaming telematics data, the framework can further be adapted to real-time and adaptive risk scoring, enabling instantaneous updating of trip-level risk as new observations arrive.

\par Taken together, these directions suggest that the proposed portfolio-anchored frequency–severity index represents an initial step toward a broader signal-based paradigm for risk quantification, with applications spanning insurance, transportation safety, and data-driven infrastructure planning.

\par\medskip\medskip\medskip\medskip\medskip\medskip\medskip\noindent\textbf{Competing interests.} The authors declare that they have no competing interests. The authors have no financial or personal relationships that could have inappropriately influenced the work reported in this manuscript.

\newpage
\appendix

\section{Classification Framework: Model, Tuning, and Evaluation}
\label{app:classification}

\par This appendix provides implementation details for the binary trip-level classification study in Section~\ref{sec:application}. We consider classes
$\mathcal{K}=\{\texttt{normal},\texttt{risky}\}$, where \texttt{normal} includes all labels that begin with \texttt{normal} (e.g., \texttt{normal1}, \texttt{normal2}) and \texttt{risky} includes the remaining labels (e.g., \texttt{aggressive}, \texttt{drowsy}). We use all $40$ trips from $6$ drivers in the UAH-DRIVESET controlled dataset. For each trip $i$, the input consists of the MLTC vector $\mathbf{N}_i=(N_{i1},\ldots,N_{iM})$ and the exposure $E_i$, defined as the number of retained coefficients after thinning. Let $y_i\in\mathcal{K}$ denote the resulting binary class label for trip $i$.

\subsection{Model, Empirical-Bayes Prior, and Scoring}
\label{app:classification:model}

\par For each class $k\in\mathcal{K}$, we use a Poisson Bayes model for the MLTC counts:
\[
N_{im}\mid (y_i=k),\lambda_{km},E_i \overset{\mathrm{ind}}{\sim} \mathrm{Poisson}(E_i\lambda_{km}),
\qquad m=1,\ldots,M,
\]
where $\lambda_{km}$ is the class-$k$ per-exposure intensity for layer $m$ and conditional independence is assumed across layers given $y_i$.

\par We place a Gamma prior on each layer intensity,
\[
\lambda_{km}\sim \mathrm{Gamma}(\alpha_{0m},\beta_{0m}),
\]
and estimate the hyperparameters $(\alpha_{0m},\beta_{0m})$ from the \emph{training set} using winsorized moment matching (with trimming rate $0.05$). The parameters $(\alpha_{0m},\beta_{0m})$ are estimated once per training set by pooling trips across both classes (i.e., the prior is shared across $\texttt{normal}$ and $\texttt{risky}$), and class separation is driven by the class-conditional likelihood aggregation below.
Specifically, for each layer $m$ we form per-exposure rates $R_{im}=N_{im}/E_i$ over the training trips, winsorize them by clipping to the empirical quantiles
$\bigl[Q_{m}(0.05),\,Q_{m}(1-0.05)\bigr]$, and compute the winsorized mean and variance
\[
\widehat{\mu}_m=\mathrm{mean}(R_{im}^{\mathrm{win}}),\qquad
\widehat{v}_m=\mathrm{var}(R_{im}^{\mathrm{win}}).
\]
We then match moments to the Gamma$(\alpha_{0m},\beta_{0m})$ parameterization,
\[
\widehat{\alpha}_{0m}=\frac{\widehat{\mu}_m^2}{\widehat{v}_m},\qquad
\widehat{\beta}_{0m}=\frac{\widehat{\mu}_m}{\widehat{v}_m}.
\]

\par Given a class-$k$ training set $\mathcal{D}_k$, conjugacy yields the posterior mean intensity used in classification:
\[
\widehat{\lambda}_{km}
=\frac{\widehat{\alpha}_{0m}+\sum_{i\in\mathcal{D}_k}N_{im}}{\widehat{\beta}_{0m}+\sum_{i\in\mathcal{D}_k}E_i}.
\]
We use an empirical class prior on each training set,
\[
\cprior_k=\Pr(y=k)\approx \frac{n_{k,\mathrm{train}}}{n_{\mathrm{train}}},
\]
where $n_{k,\mathrm{train}}$ is the number of training trips in class $k$ and $n_{\mathrm{train}}$ is the total number of training trips.

\par Let $w_m\ge 0$ be a severity weight (defined in Section~\ref{sec:frequency-model}). For a test trip $i$, we compute a severity-weighted log-score, while accounting for severity 
\begin{equation}
    D_k(i)=\log \cprior_k +\sum_{m=1}^M w_m\Bigl(N_{im}\log\widehat{\lambda}_{km}-E_i\widehat{\lambda}_{km}\Bigr),
    \label{eq:classification_log_score}
\end{equation}
and convert $\{D_k(i)\}_{k\in\mathcal{K}}$ into a risky probability by normalization:
\[
p_i=\Pr(y_i=\texttt{risky}\mid \mathbf{N}_i,E_i)
=\frac{\exp(D_{\texttt{risky}}(i))}{\exp(D_{\texttt{normal}}(i))+\exp(D_{\texttt{risky}}(i))}.
\]

\subsection{Decision Threshold and Nested Cross-Validation}
\label{app:classification:cv}

\par To identify risky trips from the probabilistic score $p_i$, we predict
\[
\widehat{y}_i=\mathbbm{1}\{p_i\ge \tau\},
\]
where $\tau\in(0,1)$ is a decision threshold. Here $p_i$ is a model-based score on $(0,1)$, but it is not guaranteed to match the true event probability in a perfectly calibrated way: depending on the fitted rates and priors from a finite training set, $p_i$ can be systematically too large or too small. Therefore, the default choice $\tau=0.5$ need not correspond to the best threshold for our data, and we choose $\tau$ data-adaptively to obtain stable classification performance.

\par Specifically, within each outer training set we select $\tau$ by maximizing balanced accuracy (BA),
\[
\mathrm{BA}(\tau)=\tfrac{1}{2}\bigl(\mathrm{TPR}(\tau)+\mathrm{TNR}(\tau)\bigr),
\]
where $\mathrm{TPR}(\tau)$ is the true positive rate (sensitivity; fraction of correctly identified risky trips among all risky trips) and $\mathrm{TNR}(\tau)$ is the true negative rate (specificity; fraction of correctly identified normal trips among all normal trips). BA provides a stable summary because it gives equal importance to detecting risky trips and correctly clearing normal trips, rather than letting one side dominate the metric. As $\tau$ varies, this criterion discourages thresholds that look good by improving only one of $\mathrm{TPR}$ or $\mathrm{TNR}$, which leads to more consistent performance across splits.

\paragraph{Inner CV for threshold selection.}
\par Given an outer-training set, we run an inner $K_{\mathrm{in}}$-fold stratified CV (we use $K_{\mathrm{in}}=4$) and compute \emph{out-of-fold} probabilities $\{p_i^{\mathrm{oof}}\}$ for all trips in the outer-training set. Specifically, for each inner fold $j=1,\ldots,K_{\mathrm{in}}$ we:
(i) fit the classifier on the inner-training subset (including re-estimation of the winsorized empirical-Bayes prior on that subset), and
(ii) predict $p_i$ on the held-out inner fold.
After looping over $j$, each outer-training trip has exactly one out-of-fold probability.

\par We then select the threshold by scanning a candidate set
\[
\mathcal{T}
=
\Bigl(\{0,\tfrac{1}{400},\ldots,1\}\ \cup\ \{p_i^{\mathrm{oof}}\}\Bigr),
\]
i.e., a uniform grid of $401$ points (step size $1/400$) augmented by the distinct out-of-fold probabilities. For each $\tau\in\mathcal{T}$ we compute $\mathrm{BA}(\tau)$ on the entire out-of-fold set and choose a maximizer; if multiple thresholds tie, we take the median of the maximizers for stability. This produces one threshold $\widehat{\tau}$ per outer split.

\paragraph{Outer CV for performance evaluation.}
\par We evaluate classification performance using either:
\begin{itemize}
\item \textbf{Repeated stratified $K_{\mathrm{out}}$-fold CV}: we use $K_{\mathrm{out}}=4$ and repeat the random stratified split $R_{\mathrm{out}}$ times (we use $R_{\mathrm{out}}=200$). In each repeat and each fold, we:
(i) fit the classifier on the outer-training set,
(ii) select $\widehat{\tau}$ via the inner procedure above (using only the outer-training data),
and (iii) predict on the outer-test fold and compute balanced accuracy. We report averages of the balanced accuracy over all folds and repeats.

\item \textbf{Leave-one-driver-out (LODO) CV}: each fold holds out all trips from one driver and trains on the remaining drivers. We include LODO because trips from the same driver share idiosyncratic driving style and are not independent; LODO directly measures generalization to an \emph{unseen driver}. Threshold selection is still nested: within each LODO training set, we run the same inner stratified CV to obtain $\widehat{\tau}$ and then evaluate on the held-out driver.
We summarize LODO performance by averaging balanced accuracy over the $6$ driver-held-out folds.
\end{itemize}

\subsection{Severity weights, $\gamma$ selection, and ablation inputs}
\label{app:classification:weights}

\par Let $\pi_m$ denote the portfolio-level membership probability of layer $m$ from the fitted severity model (Gaussian--Uniform mixture). In Section~\ref{sec:frequency-model}, we define severity weights
\[
w_m(\gamma) = \frac{\pi_m^{-\gamma}}{\sum_{\ell=1}^M \pi_\ell^{-\gamma}},\qquad \gamma>0,
\]
so that $\sum_{m=1}^M w_m(\gamma)=1$.

\par We select $\gamma$ by grid search over $\{0.1,0.2,\ldots,2.0\}$ using \textbf{LODO CV} with the \textbf{nested threshold selection} described above. For each candidate $\gamma$, we (i) form $\{w_m(\gamma)\}$ from $\{\pi_m\}$, (ii) run LODO with the threshold chosen in inner-CV-for in each fold, and (iii) record the mean balanced accuracy across the $6$ folds. We choose $\gamma^\star$ that maximizes this mean balanced accuracy. In our application, this procedure yields $\gamma^\star=1.7$. The parameter $\gamma$ controls how the severity weights are distributed across layers.

\par Finally, we compare three ablation inputs under the same classifier and the same nested-CV thresholding; equivalently, these ablations correspond to different choices of $(M,\mathbf{N}_i,\{w_m\})$ in \eqref{eq:classification_log_score}:
\begin{itemize}
\item \textbf{(A) Total frequency (single layer):} set $M=1$ and replace the MLTC vector by the scalar total count,
$N_{i1}=\sum_{m=1}^M N_{im}$, with weight $w_1\equiv 1$.
Then \eqref{eq:classification_log_score} reduces to a one-dimensional Poisson--Gamma classifier on the total tail frequency.

\item \textbf{(B) Layer-wise frequency (unweighted):} keep the full MLTC vector and set uniform weights $w_m\equiv 1/M$. Then \eqref{eq:classification_log_score} aggregates all layers equally, so depth information enters only through the separate $\widehat{\lambda}_{km}$ values.

\item \textbf{(C) Severity-weighted frequency (proposed):} keep the full MLTC vector and set $w_m=w_m(\gamma^\star)$ from the fitted severity model. Then \eqref{eq:classification_log_score} upweights rarer (deeper) layers according to portfolio-level rarity and downweights shallow layers, aligning the classifier with the proposed frequency--severity framework.
\end{itemize}
Because (A)--(C) differ only in how $\mathbf{N}_i$ and $\{w_m\}$ enter \eqref{eq:classification_log_score}, any performance gap can be attributed directly to using layer information and portfolio-based severity weighting.

\newpage
\bibliographystyle{plainnat}
\bibliography{ref}

\end{document}